\documentclass{aa}
\usepackage{graphicx}
\usepackage{psfig}
\usepackage{natbib}
\usepackage{longtable}
\usepackage{supertabular}
\def\vsini{$V\!\sin i$}

\def\omc{$\Omega/\Omega_{\rm{c}}$\ }
\def\kps{km~s$^{-1}$}

\begin{document}

\title{On the evolutionary status of Be stars.
\thanks{This study is based on observations made at ESO La Silla, Chile, OHP,
France and CASLEO.}}
\subtitle{I. Field Be stars near the Sun}
\author{J. Zorec\inst{1}
\and Y. Fr\'emat\inst{2}
\and L. Cidale\inst{3,4}
\thanks{CASLEO 
visiting astronomer, Complejo Astron\'omico El Leoncito operated under 
agreement between CONICET and National Universities of La Plata, C\'ordoba 
and San Juan, Argentina}
}

\offprints{J. Zorec}

\institute{Institut d'Astrophysique de Paris, UMR 7095 CNRS-Universit\'e 
Pierre \& Marie-Curie, 98bis Bd. Arago, 75014 Paris, France; 
\email{zorec@iap.fr}
\and
Royal Observatory of Belgium, 3 Av. Circulaire, B-1180 Bruxelles
\and
Facultad de Ciencias Astron\'omicas y Geof\'\i sicas, Universidad 
de La Plata, Paseo del Bosque S/N, 1900 La Plata, Argentina
\and Instituto de Astrof\'\i sica La Plata, CONICET, 1900 La Plata, Argentina
}

\date{--}
\abstract{A sample of 97 galactic field Be stars were studied by taking into
account the effects induced by the fast rotation on their fundamental 
parameters. All program stars were observed in the BCD spectrophotometric 
system in order to minimize the perturbations produced by the circumstellar 
environment on the spectral photospheric signatures. This is one of the first
attempts at determining stellar masses and ages by simultaneously using model 
atmospheres and evolutionary tracks, both calculated for rotating objects. The 
stellar ages ($\tau$) normalized to the respective inferred time that each 
rotating star can spend in the main sequence phase ($\tau_{\rm MS}$) reveal a 
mass-dependent trend. This trend shows that: a) there are Be stars spread 
over the whole interval $0 \la \tau/\tau_{\rm MS} \la 1$ of the main sequence
evolutionary phase; b) the distribution of points in the ($\tau/\tau_{\rm 
MS},M/M_{\odot}$) diagram indicates that in massive stars ($M \ga 12M_{\odot}$)
the Be phenomenon is present at smaller $\tau/\tau_{\rm MS}$ age ratios than 
for less massive stars ($M \la 12M_{\odot}$). This distribution can be due
to: $i$) higher mass-loss rates in massive objets, which can act to reduce the
surface fast rotation; $ii$) circulation time scales to transport angular 
momentum from the core to the surface, which are longer the lower the stellar
mass.  

}

\authorrunning{J. Zorec et al.}
\titlerunning{Evolutionary status of Be stars}

\maketitle

\section{Introduction}

 Correlations between the Balmer line emission width with \vsini\ and the 
statistical tendency of Be-type emission line profiles to be present for low 
\vsini\ values, while Be-shell type prevails at high \vsini, inspired 
\citet{S1931} model of the Be phenomenon. This model underlies the most recent 
ones. The model also assumes that there is a secularly stable B-type stellar 
critical rigid rotator \citep{1978trs..book.....T}, which builds an extended 
circumstellar envelope (CE) condensed towards the equatorial plane by 
equatorial ejection of mass. Be stars are considered to be O, B and A spectral 
type non-supergiant stars that have shown at least once some emission in the 
Balmer lines \citep{JSJ1981}. It has long been known that Be stars are fast 
rotators and that they rotate at least 1.5 to 2 times faster than B stars 
without emission \citep{1979SSRv...23..541S,Z2004}. From a homogeneous set of
\vsini\ parameters, although not corrected for effects of fast rotation,
\citet{2001A&A...378..861C} concluded that Be stars rotate on average at 
angular velocity rates $\omega =$ \omc $\sim 0.8$. \citet{1968MNRAS.140..141S}
pointed out that the \vsini\ parameters can be systematically underestimated 
if second order effects of fast rotation on absorption lines are neglected. 
Making allowance in the calculation of rotational line broadening for star 
distortion and non-uniform surface temperature in latitude 
\citep{1924MNRAS..84..665V,1924MNRAS..84..684V}, \citet{1968MNRAS.140..141S}
concluded that Be stars might actually be critical rotators. Thus, according 
to this author ``...mild prominence activity or other minor disturbances lead
to the ejection of matter..." to form the CE. These arguments were taken up 
by \citet{O2004} and \citet{2004MNRAS.350..189T}. In a study by 
\citet{FZHF2005} of rotational effects on fundamental stellar parameters it is 
shown, however, that Be stars rotate on average at $\omega \sim$ 0.9.\par
 \citet{1960MNRAS.120...33C} suggested that the Be phenomenon occurs during 
the secondary contraction phase, where the surface rotation velocity has been
spun up as a consequence of angular momentum conservation. Raw photometric 
color indices place Be stars near, or on, the TAMS (terminal-age main sequence) 
\citep{1964VeBon..70....1S, 1976ApJ...204..493S}. However, this apparent 
location in the HR diagram of some Be stars is due in part to the continuum 
emission excess produced in CE and to the over-luminosity of the central 
objects carried by the rotationally-induced stellar geometrical deformation 
and the concomitant gravitational darkening effect \citep{1924MNRAS..84..665V,
1924MNRAS..84..684V,1980ApJ...242..171S,1982A&A...109...48M,
1985ApJS...59..769S,1977ApJS...34...41C,1991ApJS...77..541C,
2004MNRAS.350..189T,FZHF2005}.\par
 If the Be phenomenon is an outgrowth of nearly critical stellar rotation, one 
of the fundamental questions becomes whether such a fast surface rotation is 
an innate property, or it is acquired at some stellar evolutionary phase. 
Two different phenomenological frames were put forward to tackle this question, 
which depend on whether Be stars are considered binaries or single stars. In 
binaries the Be phenomenon could arise after a Roche-lobe overflow event, when
one of the components gains mass and angular momentum 
\citep{1981A&A...102...17P,1987pbes.coll..339H,2000bpet.conf..668G}. While 
this mechanism cannot entirely account for the observed frequency of Be stars
\citep{1991A&A...241..419P, 1997A&A...322..116V}, it can explain Be/X-ray 
binaries \citep{2000bpet.conf..656C}. Therefore, let us assume that Be 
phenomenon concenrs only single stars only. In that case, the near critical equatorial 
velocity can either be an attribute of stars since their ZAMS (zero-age main 
sequence) phase, or a property that is acquired during their long-lived main
sequence (MS) evolutionary phase. Using moments of inertia of non-rotating 
stellar models, \citet{1970stro.coll...48H} concluded that the initial 
rotation at ZAMS must be from 1 to 4\% below the critical rate, for the 
star to become a critical rotator from core contraction in the MS. Since the
fast rotation reduces considerably the stellar momentum of inertia, according
to \citet{1982IAUS...98..299E} only 40\% under-critical rotation at the ZAMS 
would be needed to accelerate the star to the critical rotation during the MS
life span. For the star can reach critical rotation during the MS phase on
the stellar surface also depends on the initial amount and internal 
distribution of angular momentum and on its loss and further redistribution 
mechanisms \citep{1979ApJ...232..531E,2000A&A...361..101M,2000ApJ...544.1016H,
2002A&A...390..561M,2001A&A...373..555M,2003A&A...411..543M,
2002A&A...383..218S}.\par
 Apart from the mentioned photometric results, few studies deal with the 
evolutionary status of Be stars. \citet{1980A&AS...42..103J} and 
\citet{1982IAUS...98..125H} noticed a mild tendency of late Be type stars to
belong to the giant luminosity class, while early Be type stars have a 
tendency to be dwarfs as if there was a mass-age selection that underlies the 
Be phenomenon. \citet{1997A&A...318..443Z} found that the frequency of 
galactic Be stars against spectral type does not differ strongly from one 
luminosity class to another, which might suggest that the Be phenomenon can 
appear at any evolutionary stage during the MS phase. In a study of Be stars 
in open clusters, \citet{1990RMxAA..21..373F} noticed an increase of the 
frequency of these objects near the middle of the MS phase. Similarly, 
\citet{2000A&A...357..451F} concluded that the change of the frequency 
distribution of Be stars against the spectral type as a function of cluster 
age could be accounted for by assuming that the Be phenomenon occurs in the 
second half of the MS phase.\par
 Whil in the stellar count of field Be stars by \citet{1997A&A...318..443Z} 
the effects from CE-dependent over-luminosity and spectral changes due to fast
rotation were taken into account, the binning of stars by luminosity class 
groups, which was meant to represent an evolutionary-dependent separation, 
cannot be justified entirely for many stars of the sample. On the other hand, 
in \citet{1990RMxAA..21..373F} and \citet{2000A&A...357..451F}, the sampling 
of Be stars in clusters against spectral type can be incomplete. In these 
environments, Be stars are frequently detected photometrically. Since massive 
Be stars are scarce and the emission in the Balmer lines of stars cooler than 
B7 can be low, information on the appearance of the Be phenomenon in the 
relevant stellar masses may then be missing. Moreover, photometric and 
spectroscopic spectral types were not corrected for alterations due to the CE 
nor for changes induced by fast rotation. Thus, due to these shortcomings the 
conclusions drawn in those attempts are likely biased.\par
 Another way to undestand the evolutionary status of Be stars that may in 
principle solve the above inconveniences, is to study in detail a 
statistically significant number of individual Be stars, where in each star
the perturbations produced by the circumstellar emission/absorptions on the 
observed spectra are considered and account is taken of the 
rotationally-induced effects. The aim of the present paper is thus to discuss
the evolutionary stage of a well-observed sample of bright, field galactic Be 
stars, whose observational data were treated for all these deviations. The 
fact that these objects are bright enough implies that their Be character is 
well recorded. This is of particular interest for those objects either among 
the more or the less massive stars of the sample, because in general they are
not so numerous in either of these extremes of mass.\par

\section{Observational data}
 
 One of the main concerns related to the analysis of the observational data of
Be stars is to remove circumstellar emission/absorption perturbations. We 
determine the fundamental parameters of these stars using BCD 
spectrophotometric data \citep{1952AnAp...15..201C}. In this system the Balmer 
discontinuity is observed, which is characterized by two independent 
measurable quantities: the flux jump at $\lambda =$ 3700 \AA, $D$ in dex and 
the mean spectral position of the discontinuity, $\lambda_1$, presented in 
$\lambda_1\!\!-\!\!3700$ \AA. The $(\lambda_1,D)$ parameters are strong 
functions of $T_{\rm eff}$ and $\log g$ \citep{DZ1982, 1986serd.book.....Z}. 
It has been shown several times that for Be stars these two quantities are 
free of circumstellar extinction and circumstellar emission/absorption 
\citep{1991A&A...245..150Z}. Be stars may show a Balmer discontinuity with two 
separated components. While the variable component due to the CE can be, in a 
given star, either in emission or in absorption depending on the time they are 
observed \citep{1998A&AS..129..289M,1999A&A...349..151M}, the constant 
component reflects the average photospheric properties of the observed stellar 
hemisphere \citep{2002rnpp.conf..244Z}.\par
 The program stars are listed in Table \ref{tabdat}. They were observed for 
more than 50 years at OHP (France) and ESO (La Silla, Chile) with the Chalonge 
spectrograph \citep{BCD1973}, a device specially conceived to observe the 
stellar Balmer discontinuity. Since 1990 some of the program objects in the 
south hemisphere have been observed in CASLEO (San Juan, Argentina) with the 
Boller \& Chivens Cassegrain spectrograph, whose resolution in the low 
dispersion mode is similar to that of the Chalonge spectrograph 
\citep{2000ASPC..214...87C,2001A&A...368..160C} and enables one to separate 
clearly the stellar from the circumstellar Balmer discontinuity. The 
$(\lambda_1,D)$ parameters of the observed stars are listed in Table 
\ref{tabdat}(columns 2 and 3). Their r.m.s deviations are on average 
$\sigma_{\rm D} \la$ 0.005 dex and $\sigma_{\lambda_1} \la$ 0.3 \AA. \par
 The $(\lambda_1,D)$ were calibrated in the ($T_{\rm eff},\log g$) parameters
\citep{DZ1982,1986serd.book.....Z}. When these calibrations are used for fast
rotating stars, they must be considered to represent the aspect angle-averaged 
properties of the stellar photosphere in the observed stellar hemisphere. 
Hereafter we call them {\it apparent} fundamental parameters. The {\it 
apparent} ($T_{\rm eff},\log g$) sets are listed in Table \ref{tabdat}: 
$T_{\rm eff}$ and $\log g$ (columns 4 and 5). The \vsini\ parameters employed 
in the present work (column 6) are from \citet{2001A&A...378..861C} and 
Fr\'emat et al. (2005). The listed \vsini\ parameters were obtained using 
classical models of stellar atmospheres \citep{Stoeckley1973}, where the 
variation of the limb-darkening coefficient with frequency in the line is 
taken into account, so that they can be considered free of underestimations
induced by the use of constant limb-darkening coefficients 
\citep{1995ApJ...439..860C}. Nevertheless, these \vsini\ parameters must also 
be considered $apparent$.\par
\begin{figure}[t]
\centerline{\psfig{file=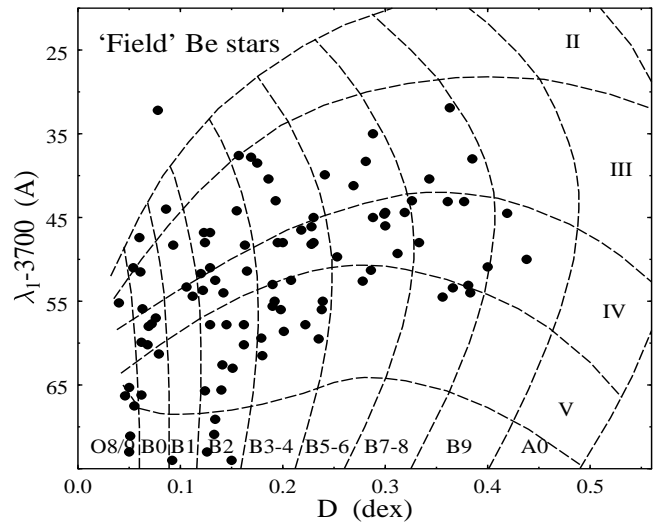,width=8.7truecm,height=7truecm}}
\caption[]{HR diagram of the program Be stars in terms of the BCD ($\lambda
_1,D$) parameters. Each curvilinear quadrilateral represents a MK spectral 
type-luminosity class group. The vertical strips demarcate the spectral types,
which are given in the abscissa. The luminosity classes corresponding to the 
horizontal strips are marked in the quadrilaterals on the right side of the 
diagram}
\label{fig1}
\end{figure}

 
 Fig. \ref{fig1} shows the HD diagram of the observed Be stars given in terms
of the observed BCD ($\lambda_1,D$) parameters. In this diagram we can see 
the tendency mentioned by \citet{1980A&AS...42..103J} and 
\citet{1982IAUS...98..125H} of late type Be stars (cooler than B5) to be on
average slightly more luminous than early Be stars. Although the 
($\lambda_1,D$) parameters of stars presented in Fig. \ref{fig1} can be 
considered free of CE perturbations, they are ``apparent" quantities because 
they still need to be treated for effects induced by fast rotation. Our 
stellar sample has a spectral type distribution that mirrors quite well that 
of the whole known Be star population near the Sun \citep{1997A&A...318..443Z}. 
The results we obtain with them can then be considered to represent fairly 
well the properties of this entire population.\par

\begin{table}[h!]
\caption{Program stars, observed and derived parameters}
\label{tabdat}
\begin{tabular}{c}
\hline
\noalign{\smallskip}
Electronic version\\
\noalign{\smallskip}
\hline
\end{tabular}
\end{table}

\section{Method}

\subsection{Stellar atmosphere models of rotating stars}

 The models of rapidly rotating stars used in the present work to describe the 
aspect angle average spectroscopic and spectrophotometric characteristics of 
early type stars are described in \citet{FZHF2005} (calculation code {\sc
fastrot}). They correspond to objects with overall rigid rotation and take 
into account their geometrical deformation as described by equipotentials in 
the Roche approximation. Allowance is also made for changes of the polar 
radius and the bo\-lo\-me\-tric luminosity produced in the stellar core. The 
lowering of the bolometric luminosity was considered to be related to the 
mass-compensation effect of rigidly rotating stellar cores 
\citep{1970A&A.....8...76S,1979ApJ...230..230C}. The non-uniform effective 
temperature distribution with latitude follows the \citet{1924MNRAS..84..665V,
1924MNRAS..84..684V} theorem as far as high enough temperatures are concerned.
For local effective temperatures lower than 8000 K we used the gravitational 
darkening calculated by \citet{1998A&A...335..647C}. The calculation code 
{\sc fastrot} e\-na\-bles us to calculate spectral lines and energy 
distributions. We can then estimate the changes produced by rotation on the 
Balmer discontinuity and the $(\lambda_1,D)$ parameters that describe it.\par

\subsection{Relation between ``$apparent$" and ``$parent$ $non-rotating$
$counterpart$" stellar parameters}

\begin{figure}[t!]
\centerline{\psfig{file=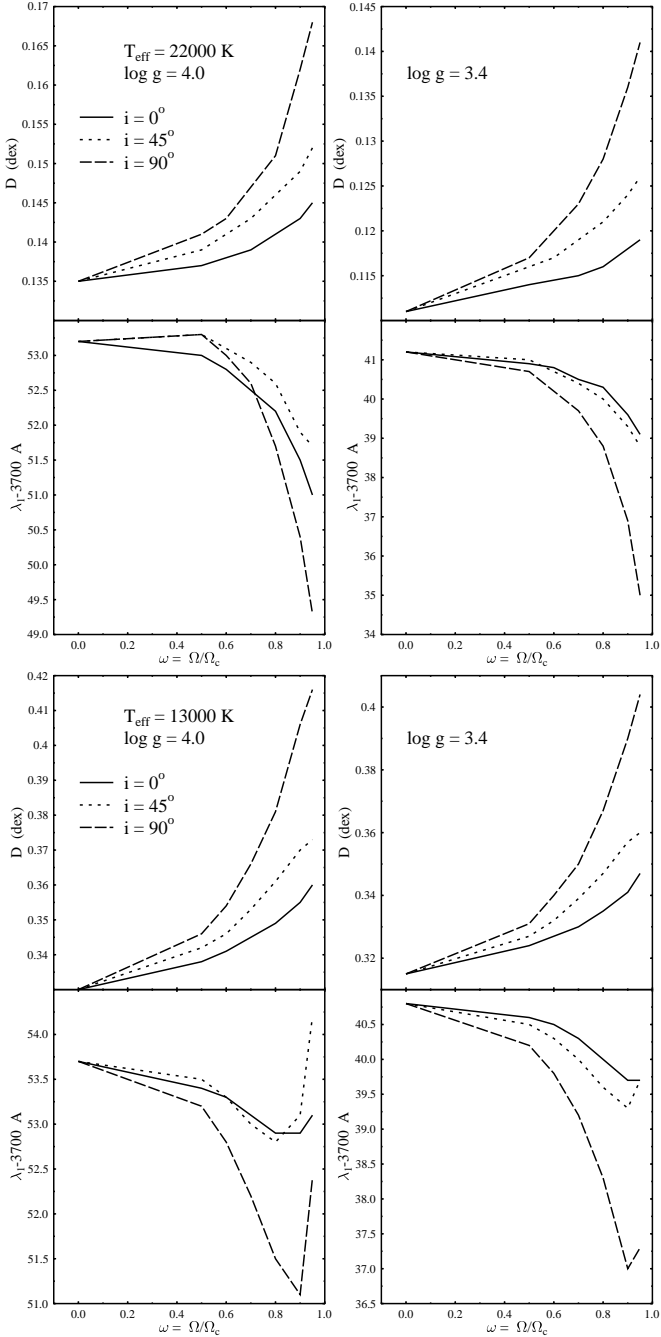,width=8.8truecm,height=18truecm}}
\caption[]{Model $(\lambda_1,D)$ parameters against the rate \omc for 
different unperturbed pairs $(T_{\rm eff},\log g)$ and several aspect 
angles~$i$}
\label{fig2}
\end{figure}

\begin{figure}[ht!]
\centerline{\psfig{file=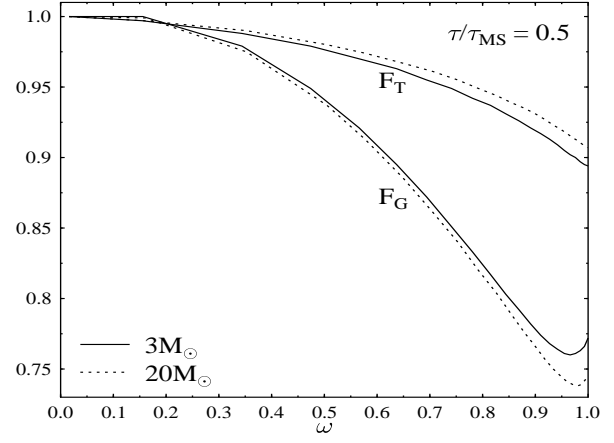,width=8truecm,height=6truecm}}
\caption[]{Functions $F_T$ and $F_G$ for $\tau/\tau_{\rm MS} =$ 0.5 and
two stellar masses}
\label{figtg}
\end{figure}

 The main purpose of the present paper is to infer stellar fundamental 
parameters that may give us some insight into the most plausible evolutionary
state of the studied Be stars, once the observed quantities are treated for 
the first order rotational effects. To derive the actual stellar mass $M$ of 
a fast rotator, we assume that the observed set of parameters 
($\lambda_1,D,V\!\sin i$) is affected by two types of rotational effects. 
There are direct changes related to the stellar geometrical deformation and 
the consequent non-uniform surface temperature and gravity distributions. 
Moreover, there are effects related to changes that the rotation produces on 
the evolution of stars. In order to take both types of effects into account, 
we proceed in two steps. Our models are built as a function of effective 
temperature and surface gravity of homologous spherical stars: the same mass, 
but without rotation. So, in a first step, from the observed ($\lambda_1,D$) 
quantities we derive the "$parent$ $non-rotating$ $counterparts$" ($pnrc$). In 
a second step, we use the {\it pnrc} ($T_{\rm eff},\log g$) sets to derive the 
($\overline{T_{\rm eff}},\log\overline{g}$) quantities, which are the 
effective temperature and gravity $averaged$ over the whole 
rotationally-deformed stellar surface. These average quantities are finally 
used as the entry parameters to the models of stellar evolution with rotation
to infer masses and ages. For consistency with other works based on the use of 
{\sc fastrot}, the nomenclature of fundamental parameters follows that adopted
in \citet{FZHF2005}.\par
 The transition from $apparent$ to $pnrc$ parameters is carried out 
considering the following transformations:
\begin{eqnarray}
\left.
\begin{array}{rcl}
        D  & = & D_o(T_{\rm eff},\log g)\times F_D(T_{\rm eff},\log 
g,\omega,i) \\
\lambda_1  & = & \lambda_1^o(T_{\rm eff},\log g)\times F_{\lambda_1}(T_{\rm 
eff},\log g,\omega,i) \\
V\sin i    & = & V_c(T_{\rm eff},\log g)\times\frac{R_e(T_{\rm eff},\log 
g,\omega)}{R_c(T_{\rm eff},\log g)}\omega\sin i \\
\end{array}
\right\},
\label{eq1}
\end{eqnarray}
\noindent where $D$ and $\lambda_1$ are the observed BCD quantities, while 
$D_o$ and $\lambda_1^o$ are those of the rotationless homologous star; $T_{\rm 
eff}$ and $\log g$ are the {\it pnrc} fundamental parameters; $i$ is the 
stellar aspect angle; $V_c$ is the critical linear equatorial velocity for 
rigid rotation; $R_c$ is the critical equatorial radius and $R_e$ the `actual'
equatorial radius at the rotational rate $\omega =$ \omc. The factors $F_D$ 
and $F_{\lambda_1}$ are functions calculated with {\sc fastrot} for rigid 
rotators, which accounts for the geometrical deformation of stars, as well as 
for the corresponding aspect angle dependent external non-uniform temperature 
and gravity distributions. Some examples of the behavior of $D$ and 
$\lambda_1$ as a function of the angular velocity rate $\omega =$ \omc and 
inclination angle $i$ for several {\it pnrc} ($T_{\rm eff},\log g$) pairs are 
shown in Fig \ref{fig2}.\par
 As in previous calculations of rotational effects on observational quantities
of B stars \citep{1970A&A.....7..120M,1972A&A....21..279M,1977ApJS...34...41C,
1991ApJS...77..541C,2004MNRAS.350..189T,FZHF2005}, the calculations of 
($\lambda_1,D)$ also show that rotational effects become conspicuous at rates
$\omega \ga$ 0.5. This limit is however lower for lower effective temperature.
The ($\lambda_1,D)$ parameters are double valued functions against $\omega$ at 
low enough {\it pnrc} effective temperatures ($T_{\rm eff} \la$ 13000 K). This 
occurs as a consequence of the gravitational darkening effect, which changes 
the average ionization balance in the stellar surface. As in stars with 
spectral types cooler than A1, the H$^-$ absorption increases progressively 
over that of b-f transitions of neutral hydrogen, which carries a decrease of 
the value of $D$. Similarly, the $\lambda_1$ folding at high $\omega$ rates 
reflects the same behavior as in late B-type giant stars, in the sense that 
pressure broadening effects on Balmer lines makes $\lambda_1$ decrease more 
rapidly the lower the effective temperature as $\log g$ decreases.\par
 Since model evolutionary tracks are presented in terms of fundamental
parameters $averaged$ over the whole stellar surface ($\overline{T_{\rm 
eff}},\log\overline{g}$), a relation between these quantities and the $pnrc$ 
$(T_{\rm eff},\log g$) determined from (\ref{eq1}) has to be used to infer 
stellar masses and ages. Nevertheless, instead of using direct $\overline{X_i}
=$ $F(X_{i,j},\omega)$ relations between $pnrc$ and $averaged$ $X$-quantities,
we prefer to iterate relations like:
\begin{eqnarray}
\left.
\begin{array}{rcl}
\overline{T_{\rm eff}} & = & T_{\rm eff}\times F_T(M,\tau,\omega) \\
\log\overline{g}  & = & \log g\times F_G(M,\tau,\omega) \\
\end{array}
\right\}
\label{eq2}
\end{eqnarray}
\noindent which are essentially of geometrical nature and where the dependency
on the stellar age $\tau$ and mass $M$ is small. Fig. \ref{figtg} shows the 
functions $F_T$ and $F_G$ for $M =$ 3 and 20$M_{\odot}$ at $\tau/\tau_{\rm MS} 
=$ 0.5.\par 
 The \vsini\ parameter on the left-hand-side of the third relation in
(\ref{eq1}) is intended to represent the $true$ rotational parameter, i.e. 
the parameter corrected for underestimations induced by the gravity darkening
\citep{1968MNRAS.140..141S,2004MNRAS.350..189T,FZHF2005}. Since 
this correction depends on the $pnrc$ $(T_{\rm eff},\log g$) and $\omega$, it 
has to be iterated simultaneously as we search for the solution of the system 
(\ref{eq1}).\par
 To solve relations (\ref{eq1}) and (\ref{eq2}) we need to specify either
the angular velocity rate $\omega$ or the inclination $i$. Since it was shown
by \citet{FZHF2005} that most Be stars rotate at $\omega =$ 
0.88$^{+0.06}_{0.04}$, which is a distribution with a very low dispersion of 
angular velocities, we can adopt $\omega =$ 0.88 to solve the relations in the 
sought parameters $M$, $\tau$ and $i$. In this paper we focus our discussion 
only on $\tau$ and $M$. We also note that the solution of (\ref{eq1}) and 
(\ref{eq2}) implies that we can translate $(\lambda_1,D)$ into $(T_{\rm 
eff},\log g)$. As empirical calibrations cannot be used, because of mixed 
rotational effects, we use model calculations. This implies that we do not 
have errors arrising from the procedure of translating $(\lambda_1,D)$ into 
$(T_{\rm eff},\log g)$, but only with those passed from the observed BCD 
quantities onto the $apparent$ fundamental parameters. The propagation of 
empirical uncertainties in the determination of $(\tau,M)$ is discussed in 
Sect. \ref{atos}.\par

\subsection{Evolutionary tracks of rotating stars}\label{evtra}

 Model tracks of stellar evolution with rotation calculated by Meynet \&
Maeder (2000, MM2000) for solar chemical composition $Z = 0.02$ were done for
different initial (or ZAMS) true equatorial rotation velocities $V_o$. In 
these models it is also assumed that in the ZAMS the stars start evolving as
rigid rotators. The use of evolutionary tracks with rotation has two 
difficulties:\par
 a) no information exists on what $V_o$ should be adopted to interpolate 
stellar masses and ages. \citet{ZLCRLFB2004} have shown that the true 
equatorial velocities of dwarf Be stars have a quite flat distribution against 
spectral type around $V \simeq$ 300 km~s$^{-1}$ ($\overline{V} \simeq $ 350 
\kps at BV0 and $\overline{V} \simeq $ 270 \kps for BV9). Calculations of 
internal angular momentum redistribution foresees that in the first 1 to 2\% 
of the MS lifetime an initial flat internal angular velocity distribution 
transforms into a step-like one, where, depending on the mass, the rotation in 
the stellar core becomes 20\% to 40\% faster than in the envelope 
\citep{1999A&A...341..181D,2000A&A...361..101M}. Since classical Be stars have
masses that range from 3 to 30$M_{\odot}$, we should then use models 
of stellar evolution calculated for somewhat higher mass-dependent initial 
velocities, ranging from $V_o \simeq$ 340 at 3$M_{\odot}$ to 420 km~s$^{-1}$ 
for 30$M_{\odot}$;\par
 b) there is some evidence for internal angular momentum redistribution in the
pre-main-sequence (PMS) evolutionary stages of stars with masses from 0.1 to 
10$M_{\odot}$ \citep{2004ApJ...601..979W} which implies that stars can start 
evolving from the ZAMS as differential rotators. If so, and depending on the 
PMS evolutionary characteristics of individual stars, the amount of 
rotational energy stored by them could be higher than the limit imposed by the 
critical rigid rotation \citep{1978trs..book.....T}. In such a case there is a
much higher mass-compensation effect on the core bolometric luminosity and 
there may be more consequences on the stellar evolution than those accounted 
for in models used in the present work.\par
 Facing the quoted unknowns on the internal rotation of the studied stars at 
the ZAMS and on their initial equatorial velocity, we estimate the effect on 
the mass and age estimates in the physical framework defined by the existing 
calculations of stellar evolution with rotation.\par
 First, we obtained stellar ages and masses with the evolutionary tracks 
without rotation for $Z = 0.02$ \citep{1992A&AS...96..269S}. Then, we derived
the same quantities using the evolutionary tracks for $V_o =$ 300 km~s$^{-1}$
and $Z = 0.02$ of MM2000. These last correspond to initial angular velocity 
rates that range from $\omega_o =$ 0.79 at $M =$ $3M_{\odot}$ to $\omega_o =$
0.52 for $M =$ $30M_{\odot}$. They produce smaller effects on the mass and age 
estimates than the slightly higher velocities $V_o=f(M)$ needed to produce the 
average main sequence $V \simeq$ 300 km~s$^{-1}$ after having undergone rapid
initial internal angular redistribution in the ZAMS. The enhanced values 
$V_o=f(M)$ imply rotational effects on stellar evolution scaled in terms of 
initial rates ranging from $\omega_o =$ 0.87 at $M =$ $3M_{\odot}$ to 0.70 for 
$30M_{\odot}$. This difference in the initial rates $\omega_o$ may have some 
significance, since rotationally-induced effects on the stellar fundamental 
parameters increase rapidly once $\omega > $0.8 and they are stronger for 
lower stellar mass \citep{FZHF2005}.\par 
 In order to estimate an order of magnitude of the effects on the estimates of 
mass and age by initial equatorial velocities larger than those used in the 
published models, we re-scaled the existing evolutionary tracks as a function 
of the abovementioned mass-dependent values $V_o=f(M)$.\par 

\subsection{Re-scaled evolutionary tracks}\label{ret}
 
 The calculations carried out by \citet{1979ApJ...232..531E} and more recently 
by \citet{2000ApJ...544.1016H,2000A&A...361..101M,2002A&A...390..561M} and
\citet{2001A&A...373..555M} show that rotation introduces several changes in 
the e\-vo\-lu\-tio\-na\-ry tracks compared to those for non-rotating stars. 
The characteristics of these changes, under the assumption that the stars 
start evolving from the ZAMS as rigid rotators, depend on: the adopted 
mass-loss rates, the initial conditions such as chemical composition and the 
initial rotational velocity $V_o$ and on the mechanisms of angular momentum
redistribution inside the star. In Be stars, the average mass-loss rate $\la 
10^{-9}$ $M_{\odot}$~yr$^{-1}$, which encompasses winds and discrete mass 
ejections, cannot lead to sensitive deviations from the evolution with the 
time-dependent variation of stellar mass already foreseen in the existing 
calculations. In this work, those changes of mass are assumed to be the same 
as that calculated for objects evolving with $V_o =$ 300 km~s$^{-1}$. However, 
depending on the initial value of the rotational velocity and the further 
phenomenon of angular momentum redistribution, there are at least three other 
outstanding changes in the evolutionary tracks of rotating stars in the MS 
phase that interest our fundamental parameter determination:\par 
 $i$) In the ($\log L/L_{\odot},T_{\rm eff}$)-plane the tracks are slightly 
shifted and rotated, so that for a given mass the starting point in the ZAMS 
is located at a lower temperature and luminosity, which reveals the 
rotationally-induced mass-compensation effect \citep{1970A&A.....8...76S};\par 
 $ii$) The MS phase is prolonged to higher luminosities than in the 
non-rotation models due to the enlargement of the H-content in the convective 
core, which is produced by the mixing processes that fuel it with fresh 
hydrogen \citep{2000ApJ...544.1016H,2000A&A...361..101M}. On the other hand,
as evolution proceeds in the MS phase, in rotating stars there is a more 
sensitive change of the moment of inertia than in non-rotating objects
\citep{1982IAUS...98..299E,2000A&A...361..101M}, which leads to an enhanced 
stretching of the star. The MS phase can then end up at lower surface-averaged 
effective temperatures than in non-rotating models;\par
 $iii$) The overall evolutionary MS life span outlasts the non-rotating case, 
on the one hand because the le\-vi\-ta\-tion effect produced by the rotation 
makes stars behave as is they had a lower effective mass 
\citep{1970A&A.....8...76S,1971ApJ...167..153B,1979ApJ...230..230C,
1982RPPh...44..831M} and on the other hand, because the mixing processes cited 
in $ii$) increase the core-burning lifetime \citep{2000ApJ...544.1016H,
2000A&A...361..101M}.\par

\begin{figure}[t!]
\centerline{\psfig{file=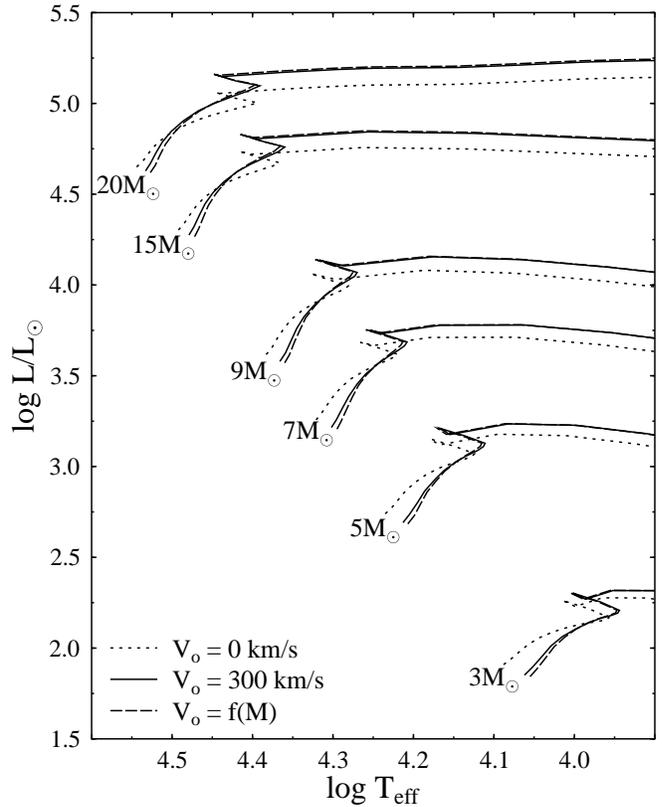,width=8.7truecm,height=11truecm}}
\caption[]{Evolutionary tracks for different initial velocities $V_o$. 
`$dotted$' lines are for evolutionary tracks with $V_o =$ 0 km~s$^{-1}$; 
`$full$' lines are for $V_o =$ 300 km~s$^{-1}$; `$dashed$' lines correspond to
$V_o =$ $f(M)$ so that $V_o =$ 340 km~s$^{-1}$ at $3M_{\odot}$ and 420 
km~s$^{-1}$ for $30M_{\odot}$}
\label{lp1}
\end{figure}

 These effects give a first order insight into the changes induced by the 
rotation, which actually encompass a series of subtle global and local 
mechanisms of mixing and angular momentum redistribution, whose incidence on
the stellar structure can be appreciated only through detailed calculations. 
In this work we use the final settings of such calculations carried out by 
MM2000 to re-scale the evolutionary tracks for initial velocities $V_o$ at 
will. We use the term `re-scale' instead of `interpolate', because some 
effects calculated in detail have been published explicitly only for a 
restricted number of stellar masses and sometimes they were typified by a 
single case that we had to generalize. Only the MS phase was treated in this 
way. The ($\log L/L_{\odot},\log T_{\rm eff}$) evolutionary paths in the 
post-MS part were constructed by translating the predictions given by 
model-evolution without rotation. Since by definition Be stars are 
non-supergiant, the supergiant region will rarely be used in the present work. 
The results of this re-scaling operation is shown in Fig. \ref{lp1}. We see 
there: 1) evolutionary tracks for non-rotating stars (dotted lines); 2) model
tracks by MM2000 calculated with $V_o =$ 300 km~s$^{-1}$ in all masses (full 
lines); 3) re-scaled tracks for a linear enhancement of the initial velocity 
$V_o$ above 300 km~s$^{-1}$ as a function of mass given by $V_o =$ 
300+[$3\times (M/M_{\odot})+30]$ km~s$^{-1}$ (dashed lines). This relation is
intended to give the initial velocities so that after the first $\approx10^4$
yr in the MS phase the Be star true equatorial velocities become $V \simeq$ 
300 km~s$^{-1}$ as observed (cf. Sect \ref{evtra}).\par
 Changes of initial velocities from $V_o =$ 300 km~s$^{-1}$ to $V_o =$ 
300+[$3\times(M/M_{\odot})+30]$ km~s$^{-1}$ seem not to produce huge effects
on the evolution path shapes. This saturation effect noted for high $V_o$ 
values was already commented on by MM2000 for hot rotating stars, where 
increased mass-loss rates due to a faster rotation partially suppress the 
effects related to an enhanced outward transport of angular momentum. Our 
`transcription' of the phenomenon in all masses for whatever $V_o$ may perhaps 
be insufficient as it depends on a simple interpolation among the few cases 
given explicitly in the literature \citep{1997A&A...322..209T,
2000A&A...361..101M}. However, we note in Fig. \ref{lp1} that: 1) the breadth 
of the MS phase of rotating models is slightly enlarged for all OB stellar 
masses as compared to the rotationless case; 2) the stellar mass that 
corresponds to a given ($\log L/L_{\odot},\log T_{\rm eff}$) point inferred in 
the lower half of the MS phase of ro\-ta\-ting objects will be slightly higher 
than the same obtained with tracks of non-rotating models; 3) stellar masses 
inferred in the upper half of the MS will not depend sensitively on the type 
of evolution model used, if masses are lower than $\sim 10M_{\odot}$, while 
for stars with $M \ga$ $10M_{\odot}$ the masses obtained from 
rotation-dependent tracks will be lower compared to those interpolated with
rotationless evolutionary paths.\par

\subsection{Stellar ages}

 The evolution time scales depend on the total angular momentum and its 
progressive internal redistribution. Let us call $\tau^o_{\rm MS}$ the time 
spent in the MS phase  by a non-rotating star (from ZAMS to TAMS) and use the 
notation $\tau_{\rm MS}$ for the MS life span of its homologous rotating 
object. Assuming that stars have rigid rotation in the ZAMS, so that their 
initial equatorial velocity is $V_o$, MM2000 have shown that between 
$\tau_{\rm MS}$ and $\tau^o_{\rm MS}$ the following relation holds:
\begin{equation}
\frac{\tau_{\rm MS}}{\tau^o_{\rm MS}} = 1.0 + \alpha
\overline{V_{\rm MS}}(M,V_o)
\label{eq3}
\end{equation}
\noindent where $\alpha =0.0013$ and $\overline{V_{\rm MS}}(M,V_o)$ in \kps is
the surface rotational ve\-lo\-ci\-ty averaged over the whole $\tau_{\rm MS}$ 
period. On account of the simplicity of relation (\ref{eq3}) and that actually
we do not have another way to scale stellar ages on evolutionary tracks for
whatever initial values of $V_o$, under the assumption that stars are rigid 
rotators in the ZAMS, let us write a similar relation to (\ref{eq3}) for the 
age $\tau(\omega)$ of a rotating object at any moment of its MS evolutionary 
phase and the respective `rest' age $\tau^o =$ $\tau(\omega\!\!=\!\!0)$:
\begin{equation}
\frac{\tau(\omega)}{\tau^o} = 1.0 + \frac{\alpha}{\tau(\omega)}\int_0^{\tau
(\omega)}V(t){\rm d}t
\label{eq4}
\end{equation}
\noindent where $V(t)$ is the time dependent surface velocity. After a rapid
decrease that lasts 1 to 2\% of the MS lifetime, in massive stars with 
mass-loss, $V(t)$ shows a roughly `linear' decrease with time (MM2000), while 
in stars with masses $M < 10M_{\odot}$, where mass-loss is negligible, it 
follows a `parabolic' increase \citep{1982IAUS...98..299E,2000A&A...361..101M}.
We can then derive the following relation between the age fraction spent in 
the MS phase by a rotating star and the homologous non-rotating object:
\begin{equation}
\frac{\tau(\omega)}{\tau_{\rm MS}} = \frac{(\tau^o/\tau^o_{\rm MS})}{1+
\gamma(\tau^o/\tau^o_{\rm MS})}
\label{eq5}
\end{equation}
\noindent in which we have :
\begin{equation}
\gamma \simeq \left[\frac{\overline{V_{\rm MS}}(M,V_o)}{V_o}-q(M,V_o)\right
]\left(\frac{\alpha V_o}{1+q(M,V_o)\alpha V_o}\right)
\label{eq6}
\end{equation}
\noindent where $q(M,V_o) =$ $V_i/V_o < 1$; $V_i$ represents the surface 
equatorial velocity after the very initial short-lasting internal angular 
momentum redistribution. Using the values of $V_i/V_o$ and $\overline{V_{\rm 
MS}}(M,V_o)$ as a function of $M$ and $V_o$ from MM2000 and 
\citet{1999A&A...341..181D} we obtained the estimates of $\gamma$ that are 
given in Table \ref{gam}. We see there that in many cases it is $|\gamma| < 
0.1$ so that to a good approximation:
\begin{equation}
\tau(\omega) \simeq \tau_{\rm MS}\left(\frac{\tau^o}{\tau^o_{\rm MS}}\right)      
\label{eq7}
\end{equation}
 On the other hand, the small values of $|\gamma|$ warrant that a very 
detailed representation of the function $V(t)$ is not relevant to evaluate
the age ratios (\ref{eq5}).\par

\begin{table}[t]
\caption{Ratios $\gamma$ as a function of mass and the initial angular 
velocity rate $\omega_o =$ $f(V_o/V_c)$}
\label{gam}
\begin{center}
\begin{tabular}{rrrr}
\hline
\noalign{\smallskip}
$M/M_{\odot}$ &$\omega_o =$ 0.5 & 0.8 & 0.9 \\
\noalign{\smallskip}
\hline
\noalign{\smallskip}
 20.0 & -0.022 & -0.008 & -0.008 \\
 15.0 & -0.011 &  0.021 &  0.029 \\
 10.0 &  0.003 &  0.056 &  0.074 \\
  5.0 &  0.022 &  0.113 &  0.154 \\
  3.0 &  0.036 &  0.170 &  0.242 \\
\noalign{\smallskip}
\hline
\end{tabular}
\end{center}
\end{table}

 Since we will be dealing with $\omega \ga$ 0.8, most Be stars studied will 
have $\tau(\omega)/\tau_{\rm MS} \la$ $ \tau^o/\tau^o_{\rm MS}$. In general
deviations from (\ref{eq7}) are small and, as expected, they are higher the 
lower the mass.\par
 For interpolation of stellar ages, we divided the ZAMS-TAMS time interval of 
each evolutionary path into 100 parts: $\tau_i$, $i=$0,1,...100, so that 
$\tau_0=$$\tau_{\rm ZAMS}$ and $\tau_{100}=$$\tau_{\rm TAMS}=$$\tau_{\rm MS}$. 
Times $\tau_i$ for the same $i$ in tracks of two consecutive masses were 
considered representing stars in homologous evolutionary stages. 
Interpolations of masses and ages were then done in the ($\log L,\log T_{\rm 
eff}$) diagrams according to this criterion of homologous evolution.\par
 A few studied objects fall in the HR-diagram strip of the secondary
contraction phase. This does not allow us to decide if these objects are still 
in the MS phase, in the secondary contraction region, or of they are already 
in the giant branch. We treated them as if they still were in the MS, since 
the short time scales involved in the remaining phases do not change much the 
estimate of the age ratios $\tau/\tau_{\rm MS}$. A small number of stars 
remain in the post-MS phase, even after all corrections of parameters for 
rotational effects. In the giant phase the evolution times were re-scaled 
using a relation similar to (\ref{eq7}).\par
 
\section{Results}
\subsection{Preliminaries}\label{pr}
 
 The major motivation of the present paper is to infer the `present-day' age 
of stars already displaying the Be phenomenon. We cannot say if the phenomenon 
has already been present in a given star for some time, or if the star will 
display it up to the end of its MS phase.\par
 As alredy noted, Fr\'emat et al. (2005) have shown that most Be stars rotate
at $\omega \simeq$ 0.88. Since this rate applies to stars that can be at 
different evolutionary stages in the MS phase, only those tracks that imply a
state of the surface velocity that fit the condition $\omega \simeq$ 0.9 at 
the required location of the star in the HR-diagram would be suitable to infer 
its mass and age. However, we do not know the individual initial velocities 
$V_o$ to build the required model tracks. This can be solved partially by 
iterating the stellar mass and its $V_o$. Nonetheless, the operation requires 
a number of subtleties that are beyond the scope of the present work and will 
be developed elsewhere. The results obtained in this section will show a 
posteriori that adopting an appropriate $V_o$ for all stars is an 
approximation that suffices for the purposes of the present work. So, in the 
present paper we calculate masses and ages adopting the models by MM2000 for 
$V_o =$ 300 km~s$^{-1}$. We estimate the magnitude of possible uncertainties 
caused by the lack of knowledge of the specific value of $V_o$. To this end, 
we determine the ($\tau/\tau_{\rm MS},M/M_{\odot}$) parameters of several HR 
`test' points in the upper and lower half of the MS phase (chosen at hoc), as 
a function of evolutionary tracks dependent on different values of $V_o$. We 
assume that the `test' points correspond to stars rotating at $\omega =$ 0.88. 
A given set of test $apparent$ parameters produce, as expected, inclination
angle-dependent series of $pnrc$ and $averaged$ fundamental parameters. 
However, the relative changes from the use of various evolutionary tracks are
the same. So, we assume that the test objects are seen at $i \simeq$ $52^o$, 
or an average inclination of rotation axes oriented at random ($52^o\simeq$ 
$\arcsin[\overline{\sin i}=\pi/4]$). For this specific angle, $apparent$ 
$X(i)$ fundamental parameters and the respective surface $averaged$ 
$\overline{X}$ obey: $X(i\simeq52^o) \sim$ $\overline{X}$ (the $X$s stand for
$\log L$, $\log g$, $\log T_{\rm eff}$, etc.).\par

\begin{table}[t!]
\caption{Comparison of masses, ages and age ratios derived from evolutionary
tracks without and with rotation}
\label{com}
\begin{tabular}{c|cc|rcc}
\hline
\noalign{\smallskip}
N$^o$&$T_{\rm eff}$&$\log g$&$M/M_{\odot}$&$\tau$ (age)&$\tau/\tau_{\rm MS}$\\
&     K       & dex    &             & years      &                    \\
\hline
1&\multicolumn{2}{l|}{Apparent parameters} & \multicolumn{3}{l}{Evolution 
without rotation} \\
\hline
\noalign{\smallskip}
&36324 & 4.043 & 24.78 & 2.24$\times10^6$ & 0.350 \\
&29119 & 3.449 & 24.03 & 5.70$\times10^6$ & 0.900 \\
\noalign{\smallskip}
&27102 & 4.100 & 11.98 & 5.53$\times10^6$ & 0.350 \\
&23529 & 3.630 & 11.93 & 1.41$\times10^7$ & 0.900 \\
\noalign{\smallskip}
&14302 & 4.157 &  4.00 & 5.72$\times10^7$ & 0.350 \\
&12169 & 3.695 &  4.00 & 1.46$\times10^8$ & 0.900 \\
\noalign{\smallskip}
\hline
2&\multicolumn{2}{l|}{$pnrc$ parameters for} & \multicolumn{3}{l}{Evolution 
without rotation} \\
&\multicolumn{2}{l|}{$\Omega/\Omega_c =$ 0.88}& \multicolumn{3}{}{}\\
\hline
\noalign{\smallskip}
&38903 & 4.185 & 27.67 & 2.68$\times10^5$  & 0.046 \\
&30941 & 3.563 & 25.10 & 5.16$\times10^6$  & 0.841 \\
\noalign{\smallskip}
&29169 & 4.234 & 13.15 & 4.23$\times10^5$ & 0.031 \\
&25291 & 3.751 & 12.67 & 1.16$\times10^7$ & 0.807 \\
\noalign{\smallskip}
&15484 & 4.280 &  4.28 & 7.37$\times10^6$ & 0.054 \\
&13176 & 3.808 &  4.19 & 1.18$\times10^8$ & 0.818 \\
\noalign{\smallskip}
\hline
3&\multicolumn{2}{l|}{Averaged parameters} & \multicolumn{3}{l}{Evolution 
without rotation} \\
&\multicolumn{2}{l|}{for $\Omega/\Omega_c =$ 0.88} & \\
\hline
\noalign{\smallskip}
&36529 & 4.037 & 25.35 & 2.24$\times10^6$ & 0.359 \\
&29026 & 3.437 & 24.24 & 5.67$\times10^6$ & 0.900 \\
\noalign{\smallskip}
&27077 & 4.089 & 12.12 & 6.29$\times10^6$ & 0.407 \\
&23478 & 3.622 & 12.15 & 1.39$\times10^7$ & 0.910 \\
\noalign{\smallskip}
&14294 & 4.149 &  4.06 & 6.62$\times10^7$ & 0.425 \\
&12163 & 3.690 &  4.09 & 1.41$\times10^8$ & 0.922 \\
\noalign{\smallskip}
\hline
\hline
4&\multicolumn{2}{l|}{Averaged parameters} & \multicolumn{3}{l}{Evolution with 
rotation} \\
&\multicolumn{2}{l|}{for $\Omega/\Omega_c =$ 0.88} & \multicolumn{3}{l}{$V_o =
300$ km/s $\forall$ masses}\\
\hline
\noalign{\smallskip}
&36545 & 4.046 & 25.88 & 2.31$\times10^6$ & 0.299 \\
&28991 & 3.411 & 22.83 & 7.60$\times10^6$ & 0.893 \\
\noalign{\smallskip}
&27089 & 4.106 & 12.61 & 5.08$\times10^6$ & 0.284 \\
&23473 & 3.614 & 11.92 & 1.64$\times10^7$ & 0.863 \\
\noalign{\smallskip}
&14294 & 4.171 &  4.27 & 4.10$\times10^7$ & 0.241 \\
&12163 & 3.690 &  4.10 & 1.59$\times10^8$ & 0.842 \\
\noalign{\smallskip}
\hline
5&\multicolumn{2}{l|}{Averaged parameters} & \multicolumn{3}{l}{Evolution with 
rotation} \\
&\multicolumn{2}{l|}{for $\Omega/\Omega_c =$ 0.88} & \multicolumn{3}{l}{$V_o =
f(M) \ga$ 300 km/s} \\
\hline
\noalign{\smallskip}
&36566 & 4.058 & 26.61 & 2.00$\times10^6$ & 0.253 \\
&28989 & 3.410 & 22.75 & 8.02$\times10^6$ & 0.898 \\
\noalign{\smallskip}
&27094 & 4.115 & 12.85 & 4.38$\times10^6$ & 0.243 \\
&23473 & 3.614 & 11.92 & 1.69$\times10^7$ & 0.867 \\
\noalign{\smallskip}
&14294 & 4.176 &  4.32 & 3.44$\times10^7$ & 0.204 \\
&12163 & 3.690 &  4.10 & 1.63$\times10^8$ & 0.846 \\
\noalign{\smallskip}
\hline
\multicolumn{6}{l}{N$^o$ = block identifier to indicate in the text the
type of}\\
\multicolumn{6}{l}{evolutionary model used}\\
\end{tabular}
\end{table}

 The results thus obtained are displayed in Table \ref{com}. The ($T_{\rm 
eff},\log g$) parameters given in columns 1 and 2 of the $1st$ block in Table
\ref{com} represent the observed, i.e. $apparent$, fundamental parameters 
which need to be treated for rotational effects. In columns 3 to 5 of the 
$1st$ block are given the masses, ages and fractions of age spent in the MS as 
reflected by models of stellar evolution without rotation 
\citep{1992A&AS...96..269S}. In columns 1 and 2 of the $2nd$ block are 
displayed the respective sets of $pnrc$ ($T_{\rm eff},\log g$) parameters of 
test points. They correspond to parameters the stars would have at rest. In
columns 3 to 4 of the $2nd$ block we display the `fictitious' quantities if 
the $pnrc$ fundamental parameters were used to derive masses and ages from 
evolutionary models without rotation. The $pnrc$ parameters ($T_{\rm eff},\log 
g$) given in columns 1 and 2 of the $2nd$ block were derived using the
\citet{FZHF2005} model atmospheres for rotating stars. Since evolutionary 
tracks of rotating stars are given in terms of fundamental parameters 
$averaged$ over the rotationally-deformed stellar surface, in columns 1 and 2 
of the $3rd$, $4th$ and $5th$ blocks we give the surface $averaged$ effective 
temperatures and gravities of the test stars rotating at $\Omega/\Omega_c 
\simeq$ 0.88. In columns 3 to 5 of the $3rd$ block are given the masses, ages 
and fractions of MS ages derived using evolutionary tracks without rotation, 
while in the same columns of the $4th$ and $5th$ blocks we give the parameters 
inferred using the original models by MM2000 with $V_o =$ 300 km~s$^{-1}$ in 
the ZAMS and the re-scaled evolutionary tracks for the $V_o =$ $f(M)$ meant to 
account for the average $V =$ 300 km~s$^{-1}$ of dwarf Be stars after the 
initial fast redistribution of the internal angular momentum (Sect. 
\ref{evtra}).\par
 The figures in Table \ref{com} reveal that for mass estimates, the 
uncertainties from possible mismatches between $V_o$ and $\Omega/\Omega_c$ at 
the required location of the star in the HR diagram are not higher than 
1$M_{\odot}$ for $M \ga 10M_{\odot}$ and they are much smaller for masses $M 
\la 10M_{\odot}$. There may be, however, strong differences in the absolute 
age estimates. These differences have to be taken into account when 
comparisons must be done with ages of stars in environments like clusters that 
were inferred from non- or slowly-rotating stars. Fortunately, the age 
fractions $\tau/\tau_{\rm MS}$ are much less sensitive to detailed 
calculations of stellar evolution. From the $4th$ and $5th$ block we see that 
the choice of tracks with rotation results in higher uncertainties on the 
fractions $\tau/\tau_{\rm MS}$ in the first evolutionary stages of the MS 
phase than the end of this phase.\par
 We can then conclude that mass determinations are not strongly sensitive to 
the type of evolutionary track used. On the other hand, the use of models for 
rotating stars that take into account the average rotational charactersitics 
of fast rotators in the dwarf state of the MS leads to estimates of fractional
ages which are not sensitive to the specific initial value of $V_o$ around 300 
km~s$^{-1}$. This ensures that the models used in the present work lead to
reliable statistical insights on global distributions and possible 
mass-dependencies of fractional ages at which the Be phenomenon occurs. It 
could be, however, suitable to proceed to more detailed iterations when 
discussing individual objects whose absolute ages are to be determined.\par

\begin{figure*}[t!]
\centerline{\psfig{file=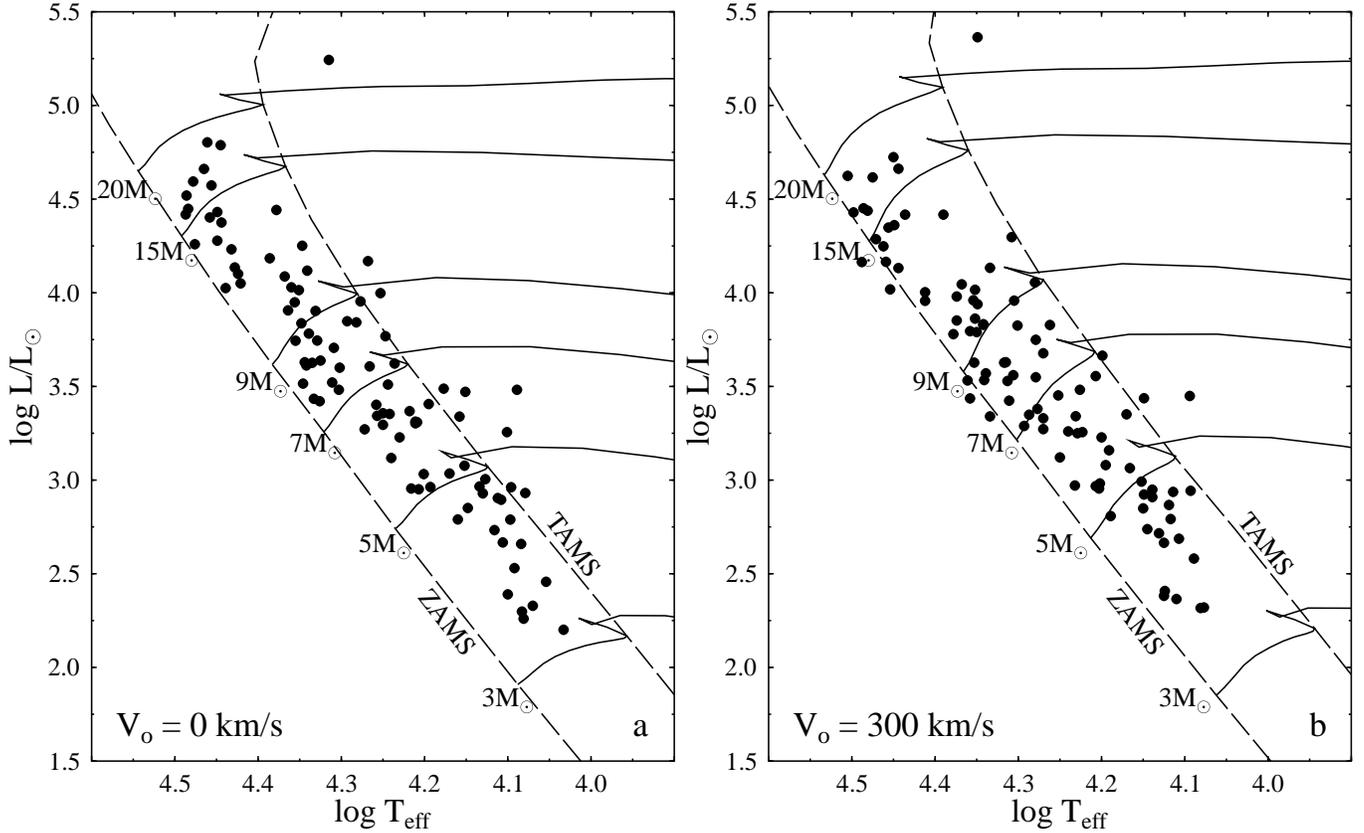}}
\caption[]{HR diagrams of the studied Be stars. a) Apparent ($\log L/L_{\odot
},\log T_{\rm eff}$) with evolutionary tracks without rotation. b) Surface 
averaged ($\log L/L_{\odot},\log T_{\rm eff}$) and evolutionary tracks with 
initial an equatorial rotation velocity $V_o =$ 300 km~s$^{-1}$ for all 
masses.}
\label{fighrs}
\end{figure*}
 
\subsection{Application to observed stars}\label{atos}

 The entry parameters used to derive stellar masses and ages using relations
(\ref{eq1}) and (\ref{eq2}) were considered with their $1\sigma$ uncertainty 
bars: $X =$ $X_o\!\pm\!\sigma_x$ ($X_o$ stands for surface averaged $\log L$ 
or $\log g$ and $\log T_{\rm eff}$). Each interval 
($X_o\!\!-\!\!\sigma_x,X_o\!\!+\!\!\sigma_x$) was divided into 7 parts, so 
that the solutions of relations (\ref{eq1})-(\ref{eq2}) and for each star 
interpolations in the HR diagrams were performed for all possible combinations 
of individual sub-$X_i$ entry parameters. Hence, for each star we obtained 
$8^2$ solutions that determined the respectively $\tau$- and $M$-distributions 
of the solutions (most of them are not symmetrical). From these distributions 
we adopted the {\it modes} as the most probable results, as well as the 
corresponding average $1\sigma$ dispersion, to account for the related 
uncertainties. We note that the uncertainties affecting the $apparent$ 
fundamental parameters are those from the observed ($\lambda_1,D$) quantities.
We also have the uncertainty of ($+0.06,-0.04$) around the adopted rotation 
rate $\omega =$ 0.88 that could affect the results. Nevertheless, the global 
changes that will imply on the ($\tau/\tau_{\rm MS},M/M_{\odot}$) diagram the 
treatment of the fundamental parameters with $\omega = 0$ or $\omega = 0.88$, 
justify neglecting the small dispersion $\delta\omega =+0.06,-0.04$. The 
$apparent$ and surface $average$ ($T_{\rm eff},\log g$) sets are given in 
Table 1. In this table we also reproduce the obtained ages, masses and MS age
fractions derived using evolutionary models without and with rotation.\par
 Fig. \ref{fighrs}a) shows the HR diagram of the studied stars in terms of 
their apparent ($\log L,\log T_{\rm eff}$) parameters and where are shown also 
the evolutionary tracks for non-rotating stars \citep{1992A&AS...96..269S}.
Fig. \ref{fighrs}b) shows the HR diagram of the same stars, but in terms of 
their surface $averaged$ ($\log\overline{L/L_{\odot}},\log\overline{T_{\rm 
eff}}$) quantities, where we assumed all stars rotate at $\omega =$ 0.88. In 
this figure are also shown the evolutionary tracks for rotating objects that 
start evolving from the ZAMS as rigid rotators with equatorial velocity $V_o 
=$ 300 km~s$^{-1}$ (MM2000).\par

\begin{figure*}[t]
\centerline{\psfig{file=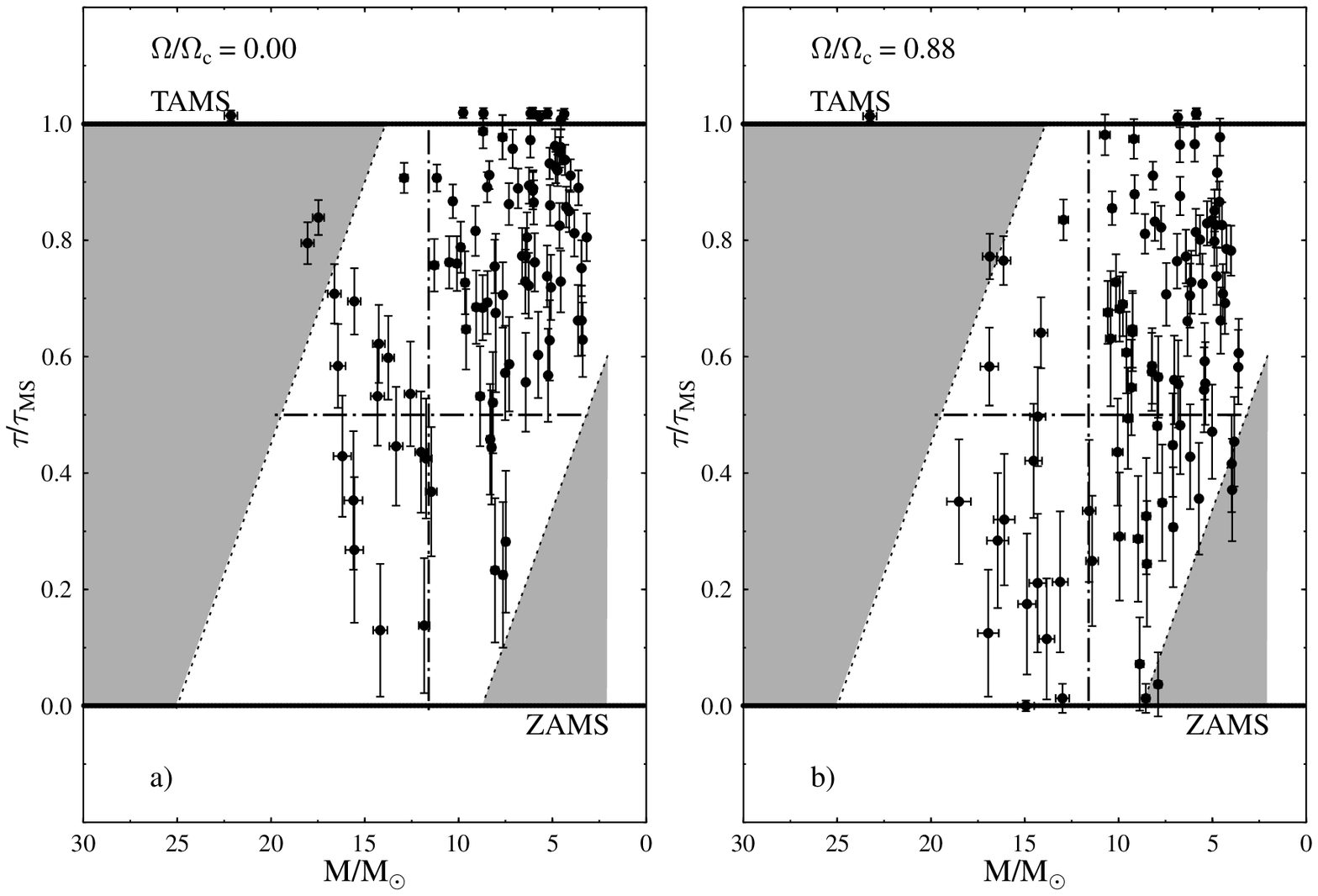}}
\caption[]{Age ratios $\tau/\tau_{\rm MS}$ of the studied Be stars against 
the mass. a) Parameters derived using evolutionary tracks without rotation;
b) parameters derived using evolutionary tracks with rotation}
\label{agem}
\end{figure*}

 In spite of the widened MS phase of rotating stars and corrections made to 
their fundamental parameters for rotational effects, four program stars lie
in the post-MS `bright giant' region. They are HD23630, HD45910, HD183656 and
HD217675, whose true \vsini\ in km~s$^{-1}$ and estimated inclinations $i$ are 
(149;$45^o$), (254;$76^o$), (274;$77^o$) and (272;$90^o$) respectively. These
objects are referenced in the literature as binaries or multiple systems. 
Their apparently too `evolved' character could be due to a merging effect of 
fundamental parameters from several components.\par
 The translation of the HR positions in Fig. \ref{fighrs} into masses 
$M/M_{\odot}$ and age fractions $\tau/\tau_{\rm MS}$ is given in Table 1 and 
in Fig. \ref{agem}. Fig. \ref{agem}a) shows the distribution of points ($\tau
/\tau_{\rm MS},M/M_{\odot}$) obtained for the program stars using the original
or $apparent$ fundamental parameters and the evolutionary tracks without 
rotation \citep{1992A&AS...96..269S}. The plotted error bars correspond to the
$1\sigma$ dispersions of the $8^2$ $\tau/\tau_{\rm MS}$- and $M/M_{\odot}$-
solution distributions. Fig. \ref{agem}b) shows the same type of distribution, 
but where parameters were obtained using surface $averaged$ fundamental 
parameters for $\omega =$ 0.88 and models of stellar evolution with rotation 
calculated by MM2000 for $V_o =$ 300 km~s$^{-1}$.\par
 We note that the uncertainty bars affecting the mass determinations are small,
while those concerning the MS life fractions are in most cases quite large.
This indicates that errors in the determined quantities due to approximate
choices of $V_o$ can be concealed within the uncertainties related to the 
observations.\par
 In both diagrams of Fig. \ref{agem}, points are spread over the whole interval 
of age fractions $0.0 \la$ $\tau/\tau_{\rm MS} \la 1.0$, which suggest that
the Be phenomenon may appear at any stage of the stellar evolution in the MS 
evolution phase. There is, however, a difference between solutions where 
rotation is taken into account and those where it is not. If we were not aware
of rotational effects, Fig. \ref{agem}a) would suggest that 86\% of stars are 
above the $\tau/\tau_{\rm MS} =$ 0.5 limit. Fig. \ref{agem}b) shows, however, 
that when the fast rotation of Be stars is taken into account, the fraction of 
stars in our sample above $\tau \simeq$ $0.5\tau_{\rm MS}$ drops to 62\%.\par
 Another important result appears when we separate the stars into $massive$ 
($M \ga 12M_{\odot}$) and $less\ massive$ ones ($M \la 12M_{\odot}$). We see 
then that in Fig. \ref{agem}b) that the stars are distributed as follows:\par 
\begin{table}[h!]
\begin{center}
\begin{tabular}{rcc|c}
  [upper MS half] & &     30\%            &       65\%           \\
\cline{1-1}\cline{3-4}
  [lower MS half] & &     70\%            &       35\%           \\
\noalign{\smallskip}
           & & $M \ga 12M_{\odot}$ & $M \la 12M_{\odot}$  \\
\end{tabular}
\end{center}
\end{table}
\noindent The mass-dependent division of Be stars regarding their position 
with respect to $\tau =$ $0.5\tau_{\rm MS}$ implies that {\it the Be 
phenomenon in massive stars tends to appear on average at smaller $\tau/\tau
_{\rm MS}$ age fractions than in the less massive stars}. The same phenomenon 
is also suggested in Fig. \ref{agem}b) by the stars with $M \la 12M_{\odot}$ 
in the lower MS half region, as they all lie above a diagonal that starts 
among the more massive objects at $\tau/\tau_{\rm MS} =$ 0 and ends at 
$\tau/\tau_{\rm MS} \simeq$ 0.5 for $M \simeq 3M_{\odot}$. Since dwarf
Be stars rotate on average with $V \simeq$ 300 km~s$^{-1}$ \citep{ZLCRLFB2004}, 
either we adopted $V_o =$ 300 km~s$^{-1}$ or $V_o =$ $f(M) \ga$ 300 
km~s$^{-1}$. It can be shown that the initial angular velocity rate $\omega_o$ 
is higher the lower the stellar mass (cf. Sect. \ref{evtra}). The fact that in 
the $M \la 12M_{\odot}$ region stars are above this well defined 
mass-dependent slope might suggest that the mechanisms of angular momentum 
transport that accelerate the stellar surface up to a near/or critical 
velocity have relatively longer time scales as the stellar masse 
decreases.\par
 We also obtained masses and age ratios $\tau/\tau_{\rm MS}$ with the 
re-scaled evolutionary tracks for $V_o =$ $f(M) \ga$ 300 km~s$^{-1}$ (cf. 
Sect. \ref{ret}). The tilted character of the trend of points obtained is the 
same as the one presented in Fig. \ref{agem}b). There is, however, a slightly
more pronounced downard drift of points.\par
 In Fig. \ref{agem} we notice that 72\% of the studied objects have masses 
from 3 to 12$M_{\odot}$. On the other hand, there is a striking lack of 
massive Be stars with $M \ga 16M_{\odot}$. This lack can be due to several 
reasons: 1) smaller IMF for massive objects; 2) the fast evolution of massive
stars, so that many of them in the solar neighbouring regions could have 
alreday migrated towards the supergiant phase; 3) the CE can be blown away by
the radiation pressure of the `potential' hottest Be stars 
\citep{1975PASP...87..777M}.\par 
 
\section{Discussion}

\subsection{Comparison with previous works}

 The HR diagram drawn in terms of the $apparent$ BCD ($\lambda_1,D$) 
parameters confirms the tendency noticed by \citet{1980A&AS...42..103J} and 
\citet{1982IAUS...98..125H} that late-type Be stars can be on average slightly 
more luminous than early Be stars. This result can also be likened to the 
trend shown in Fig. \ref{agem}a) which concerns the apparent stellar 
parameters. \par 
 Our finding of the Be phenomenon appearing at any evolutionary phase is in 
agreement with similar assertions by \citet{1982A&A...109...48M} and 
\citet{1985ApJS...59..769S} based on studies of Be stars in clusters. However,
the scatter of points over the whole MS life span found by these authors can
in part be due to the photometric data that were not corrected for CE 
perturbing effects and changes introduced by the fast rotation.\par
\citet{1997A&A...318..443Z} have concluded that the frequency of Be stars as 
a function of spectral type is roughly the same in all luminosity classes. 
These authors have dealt with a sample of field Be stars three times larger 
than the sample studied in the present work. However, they have gathered the 
studied objects in three wide groups of luminosity classes, where both 
evolutionary and rotational effects on the apparent luminosity class were 
mixed. Statistical corrections for CE flux excesses and for changes related to
fast rotation were introduced. Since these corrections focused on the 
countings per group of luminosity class separately, they concerned only 
absolute magnitudes and spectral types. Moreover, in the number frequencies 
N(Be)/N(B+Be) against spectral type, not only do B stars without emission 
greatly outnumber Be stars, but they represent a highly heterogeneous group in 
physical characteristics. Their evolution then has different characteristics 
to that of Be stars, which is thus a source of deviations in the count 
frequencies that cannot be ascribed to Be stars. Hence, the result found by 
\citet{1997A&A...318..443Z} should be considered as intermediate to those 
shown in Fig. \ref{agem}a) and Fig. \ref{agem}b).\par
 When it comes to  Be stars in clusters, statistics can be incomplete in the 
hotter and in the cooler extremes of spectral types. The scarcity of massive 
Be stars was already discussed in Sect. \ref{atos}. To this add the possible 
exclusion of genuine massive Be stars in young clusters. In the accretion 
pre-main-sequence (PMS) star formation paradigm \citep{1993ApJ...418..414P,
1994ApJS...95..517B}, stars with masses $M \ga$ $10M_{\odot}$
\citep{2001A&A...373..190B,2002ASPC..267..179M} may have a period of MS life 
when they can still be imbedded in the protostellar nebula. At these early
evolutionary stages, there may be genuine fast rotators that have already 
formed, by mass-loss processes, some circumstellar disc. Thus, their observed
emission has two sources, the disc and the leftovers from the accretion 
nebula. In spite of their mixed Be and Herbig AeBe character, these objects 
should not then be discarded from the Be statistics. For late type Be stars, 
their number is reduced because the low effective temperature maintains 
hydrogen atoms almost neutral, thus disabling possible existing CE to produce 
observable Balmer emission. Since the existing studies of frequencies of Be 
stars in clusters have not taken into account rotationally-induced effects in 
the spectral classifications, they should be compared with our results 
presented in Fig. \ref{agem}a). Thus, if in this figure we disregard Be stars 
with masses $M \ga 12M_{\odot}$ and $M \la 5M_{\odot}$ to mimic possible 
misdetections, we can see that the stars in the upper half of the diagrams 
widely outnumber those in the lower half. This readily accounts for the 
suggestion made by \citet{2000A&A...357..451F} that the Be phenomenon tends to 
appear in the second half of the MS life time.\par

\subsection{Tilted distribution of points}

 Let us discuss briefly the tilted trend of points shown in Fig. \ref{agem}b). 
 It is known that the mass-loss phenomenon in Be stars has two main components:
a) winds with rates of the order of $\dot M \sim$ $10^{-9}M_{\odot}{\rm 
yr^{-1}}$; b) discrete mass-loss events of $\delta M \sim$ $10^{-9}M_{\odot}$
underlying light outbursts several times per year \citep{2000bpet.conf..348H,
Z2004}. Be stars with masses $M \la 12M_{\odot}$ spend some $10^{7}$ to 
$10^{8}$ yr in the MS phase (MM2000), so they can be considered as evolving at 
a nearly constant mass regime. Although the mass-loss rates currently assumed
for stars in the $12 \la M/M_{\odot} \la 25$ mass interval are higher than 
quoted above, they cannot account for a total loss of the order of $\sim 
10M_{\odot}$ during the MS phase to explain the slope of the upper edge of the 
strip of points in Fig. \ref{agem} in terms of a drift towards the less 
massive side as the stars evolve from ZAMS to TAMS.\par 
 The right-hand slope of the lower edge of the trend in Fig. \ref{agem}b) may
suggest that B stars of low mass need to spend some mass-dependent time in the 
MS before they can display the Be phenomenon; i.e. the time needed for the 
surface velocity be spun up to \omc\ $\sim 0.9$.\par
 We can also speculate on the tilted aspect of the distribution of points in
Fig. \ref{agem}b) as produced by a mass-dependent distribution of initial 
equatorial velocities in the ZAMS, i.e. that massive stars start their MS 
phase with higher average rotational velocities than the less massive ones 
relative to the respective critical velocities. While the suggestion of a 
mass-dependent evolution of surface rotational velocities needs a thorough 
theoretical study, some clues on the initial rotation in the ZAMS can be 
obtained from the basics of stellar structure and correlations drawn from 
observations.\par
 Evolutionary models of rotating stars \citep{2000A&A...361..101M,
2002A&A...390..561M} suggest that we can divide the interior of stars in the
ZAMS roughly into two regions, $core$ and $envelope$, which rotate each at 
different, but respectively at near uniform angular velocities. The total 
stellar angular momentum $J$ can then be written as:
\begin{equation}
\left.\begin{array}{lcl}
J & = & k_{\rm co}M_{\rm co}R_{\rm co}^2\Omega_{\rm co}+k_{\rm E}(M_*-M_{\rm 
co})(R^2-R_{\rm co}^2)\Omega_{\rm E}     \\
  & = & k_{\rm E}M_*R^2\Omega_{\rm E}\times f   \\
\end{array}
\right\}
\label{eq8}
\end{equation}
\noindent where the $k_{\rm co}$ and $k_{\rm E}$ are the gyration radii; 
$M_{\rm co}$ and $M_{\rm E}$ are the masses; $R_{\rm co}$ and $R_{\rm E}$ the
radii; $\Omega_{\rm co}$ and $\Omega_{\rm E}$ are the angular velocities of 
the $core$ and $envelope$ respectively; '*' stands for the whole star. To 
obtain from (\ref{eq8}) an insight on the mass dependence of the initial 
velocity $V_o$, let us note that in MM2000 models we have $M_{\rm co}/M_* \la
0.4$, $R_{\rm co}/M_* \la 0.4$, $\Omega_{\rm co}/\Omega_{\rm E} \la 3$. 
Moreover, assuming that $k_{\rm co} \sim$ 0.132 and $k_{\rm e} \sim$ 0.04 as 
it would correspond to `average' polytropes of indices $n = 1.5$ for the 
convective core and $n = 3.0$ for the radiative envelope, we obtain that 
$f$ in relation (\ref{eq8}) is fairly independent of mass and $f \sim 1$.\par
 On the other hand, using observed \vsini\ values \citet{1987PASP...99.1322K} 
has shown that on average the angular momentum of dwarfs depends on the 
stellar mass as:
\begin{equation}
J \sim M_*^{2.0}.
\label{eq9}
\end{equation} 
 Since the initial MS radii of stars $R_o$ scale with the stellar mass as:
\begin{equation}
R_o \sim M_*^{0.6} \ \ \ \ \ (M_* \ga 2M_{\odot}),
\label{eq10}
\end{equation} 
\noindent from (\ref{eq9}) for $R = R_o$ the ratio of the average initial 
equatorial velocity of stars $V_o = \Omega_{\rm E}R_o$ normalized to the 
critical velocity $V_{\rm cr} \simeq (GM_*/1.5R_o)^{1/2}$ is related to the 
stellar mass as:
\begin{equation}
\frac{V_o}{V_{\rm cr}}\sim \frac{M_*^{0.2}}{k_{\rm E}}
\label{eq11}
\end{equation} 
 Knowing that for the whole range of stellar masses concerned by the Be 
phenomenon, models of the stellar interior in the ZAMS are homologous, it 
follows that the gyration radius $k_{\rm E}$ is independent of mass over the 
surveyed range of masses. Thus, (\ref{eq11}) gives us a first insight into the 
possible initial dependence of rotational velocities on mass, which implies 
that stars with masses $M_* \ga 15M_{\odot}$ can have initial rotational 
velocities 25\% closer to the critical one than those with masses $M_* \la 
5M_{\odot}$. This mass dependence is small indeed. However, to account entirely 
for the noted tendency that hot stars reach the near/or critical rotation at 
smaller ratios $\tau/\tau_{\rm MS}$ than the cooler ones, there must still be 
some interaction between rotationally and evolutionary-induced changes on the
stellar moment of inertia that can be felt as due to a small mass-related 
decrease of the gyration radius.\par 
 Let us finally note that if we neglect the possible small mass-dependence of
$k_{\rm E}$ from $M_* \sim 15M_{\odot}$ to $ \sim 5M_{\odot}$, relations 
(\ref{eq10}) and (\ref{eq11}) will imply that $V_o(15)/V_o(5) \sim$ 1.6, while 
from observations of dwarf Be stars, which are on average at a later 
evolutionary stage than implied by $V_o$, we derive $V(15)/V(5) \sim$ 1.3 (cf.
Sect. \ref{evtra}). This suggests that to obtain results of the kind shown in 
Fig. \ref{agem}b), it would be better to use models with $V_o =$ $f(M)
>$ 300 km~s$^{-1}$ (cf. Sect. \ref{ret}), rather than $V_o =$ 300 km~s$^{-1}$.
Such a choice would imply the use of re-scaled evolutionary tracks. We 
preferred, however, to keep the original models by MM2000, since they depict 
the effects of fast rotation in a more consistent way and because altogether 
they give reliable orders of magnitude of these effects.\par 
 
\subsection{Comparison with model predictions}

 The noticeable lack of Be stars with masses $M \ga$ $12M_{\odot}$ in the 
upper half of the MS can be explained as the natural result of angular 
momentum loss produced by mass-loss, whose rate increases with the stellar 
mass. This loss reduces the surface equatorial rotational velocity 
con\-ver\-ting the star into a much lower rotating object ($\omega \ll 0.9$).
Conversely, the increase of the number of Be stars with masses $M \la$ 
$10M_{\odot}$ in the upper half of the MS must be an effect of internal 
coupling, where the angular momentum is conveyed from the stellar core to the
surface by the meridional circulation. The time scale of the me\-ri\-dian 
circulation is roughly $\tau_{\rm circ} \sim$ $\tau_{\rm KH}/\eta$ where 
$\tau_{\rm KH}$ is the Kelvin-Helmholtz time and $\eta$ is the ratio of the 
centrifugal force to the gravity. Since for Be stars it is $\eta \sim$ 1, 
$\tau_{\rm circ}$ ranges with the stellar mass as $\tau_{\rm circ} \sim$ 
$4\times 10^6(M/M_{\odot})^{-2}$ yr. It then becomes clear that the lapse of
time to reach $\Omega/\Omega_{\rm c} \sim 1$ is longer the smaller the stellar
mass (MM2000 and Maeder 2004, private communication), similar to model 
predictions.\par
 
\subsection{On the rotational velocities and star formation regions}

 From {\sc hipparcos} parallaxes we see that 84\% of stars in our sample lie 
in a region within 500 pc of the Sun, of which 62\% are within 300 pc. It is 
also noteworthy that 45\% of the more massive objects of this sample, hotter 
than 22000 K, are within 300 pc and 35\% are between 300 and 500 pc. Only two 
stars are at $d \sim$ 800 pc. The stellar sample studied should not then 
be characterized by strong differences in the initial metallicity. In fact, 
low metallicities might favor the fast rotation in some cases 
\citep{1999A&A...346..459M}, but it cannot be the case for the massive Be 
stars of our sample. According to \citet{2002A&A...390..561M} low metallicity
reduces the mass-loss rate, which favors the conservation of angular momentum 
in the stellar surface and so, the existence of higher surface rotational 
velocities. Also, \citet{2001A&A...373..555M} predict that the increase of the 
$\Omega/\Omega_{\rm c}$ ratio towards 1 in massive stars is faster in mo\-dels
with low metallicity. It would then be interesting to test this prediction by 
obtaining diagrams like those in Fig. \ref{agem} for Be stars in environments 
with quite different metal abundance, an aim that we will pursue in subsequent
work.\par

\subsection{On the initial conditions}

 The calculated properties of an evolving object with rotation in the MS 
depend on the assumed initial conditions. The results given in the present 
work depend on predictions made for stars which began evolving in the ZAMS
as rigid rotators. This choice may, however, be not the only possible. 
Similarly to ZAMS, which is likely a computational `landmark'
\citep{1981ApJ...243..625E}, rigid rotation in the ZAMS may be a 
simplification too. On the one hand, the necessity and/or definition of a ZAMS
for massive stars is not clear, as it happens in the accretion paradigm of 
star formation \citep{1993ApJ...418..414P,1994ApJS...95..517B,
2002ASPC..267..179M}. On the other hand, due to hydrodynamical instabilities,
an initial rigid rotation switches rapidly (some $10^4$ yr) into a 
differential rotation \citep{1999A&A...341..181D,2000A&A...361..101M}. Such a
differential rotation may then be present before the `ZAMS' phase.\par
 In the classical PMS evolution frame, based on the contraction of a constant
mass sphe\-re until the gravitational energy release increases the central 
temperature enough to trigger the nuclear reactions, rigid rotation in the 
ZAMS was generally justified because: a) dynamical stability against 
axisymmetric perturbations could be warranted for rigid stellar rotators 
\citep{1987A&A...176...53F}; b) it was assumed that in the PMS the full 
convection phase stars become rigid rotators. However, recent 2-D 
hydrodynamical calculations show that convection does not maintain rigid 
rotation, but it rather produces an internal angular velocity distribution 
profile $\Omega(\varpi)\propto$ $\varpi^{-p}$ ($\varpi =$ distance to the 
rotation axis), which is intermediate between complete redistribution of 
specific angular momentum $(p=2)$ and rigid rotation $(p=0)$ 
\citep{1998ApJ...499..340D,2000ApJ...543..395D,2001ApJ...552..268D}. 
Furthermore, in the accretion formation scheme the star is a mass and angular
momentum gaining object \citep{1993ApJ...418..414P,1984ApJ...279..363M,
1984ApJ...286..529T}. Meanwhile a non-rotating star gains mass, it first 
undergoes a full-convection period, then a radiative core is developed, the 
star swells until a temporary full radiative state is attained and finally the 
core becomes convective. During this period the forming star stretches and 
contracts, but it can also undergo magnetic interaction with the accretion 
disc \citep{2002A&A...383..218S}. These phenomena can produce an uneven 
distribution of the angular momentum inside the star. There is also some 
decoupling of stellar internal regions, since the time scales of angular 
momentum redistribution in convective and radiative zones are different 
\citep{1981ApJ...243..625E}. If the star acquires mass through an accretion 
disc, which is probably in Keplerian rotation, a huge gain of angular must 
take place \citep{1981A&A...102...17P}:

\begin{equation}
\dot{J}(t) \approx [GR(t)M(t)]^{1/2}\dot{M}(t)
\label{gainj}
\end{equation}
 
\noindent where $G$ is the gravitional constant, $R$ and $M$ are the 
time-dependent radius and mass of the star respectively and $\dot{M}$ is the 
mass accretion rate [$\dot{M}\sim$ $10^{-5}(M/M_{\odot})^{\phi}$ $M_{\odot
}$/yr with ${\phi}\sim$ 1.0-1.5 \citep{2002ASPC..267..179M}]. Since the rigid
rotation implies a very low amount of rotational energy, it cannot be then 
excluded that a considerable load of angular momentum may exist somewhere deep 
inside the star, so that an a priori assumption of rigid rotation can be 
difficult to justify. However, it was noticed by \citet{1999A&A...349..189S,
2002A&A...381..923S} that a powerful enough differential rotation can create 
magnetic fields. Different regions inside the star could then be ``locked" 
to recover some degree of rigid rotation \citep{2003A&A...411..543M}.\par 
 \citet{2002A&A...383..218S} has suggested that given an appropriate range of
surface magnetic fields, stars may gain angular momentum through an effective
magnetic accretion in the PMS phase. \citet{2002A&A...383..218S} notes that 
the rotation can be faster the more massive the star, because the 
in\-te\-rac\-tion with the circumstellar matter lasts less time. Nevertheless, 
the final balance between losses and gains of angular momentum produced by 
interactions with the circumstellar environment were not definitively 
established, nor were its consequences on the internal rotation law of the 
star. In particular, very little was said about the amount of rotational 
kinetic energy the star is left with after these interactions.\par
 Rigid rotation puts an upper limit onto the amount of rotational kinetic 
energy ${\cal K}$ a star can store. At rigid critical rotation, an early-type
star has ${\cal E}_c=$ ${\cal K}_c/|{\cal W}|$ $\simeq$ 0.015 (${\cal W}$ = 
gravitational potential energy). In a star with internal differential rotation,
the same surface rotations may correspond to higher values of ${\cal E}$, 
which may then carry stronger stellar deformations, gravity darkening effects 
and internal hydrodynamical instabilities. For an order of magnitude estimate, 
Table \ref{conv} gives the rotational kinetic energy ratios ${\cal K}(p)/{\cal 
K}(\!p\!\!=\!\!0) $ and energy ratios ${\cal E}(p)=$ ${\cal K}(p)/|{\cal 
W}(p)|$ for an internal rotational law $\Omega(\varpi) \propto$ $\varpi^{-p}$, 
assuming that the stellar surface rotates at $\Omega_{\rm s}/\Omega_{\rm c} =$ 
0.9 as occurs on average for Be stars. These values were obtained using two 
dimentional models of stellar structure \citep{1988CRASM.306.1265Z}. We see 
that for a mild differential rotation $p = 0.4$ it is ${\cal E} >$ ${\cal 
E}_c$ and that for $p=0.7$, an average rotation law set by convection, ${\cal
E}$ is nearly twice as high as for a critical rigid rotator. In such a case 
there is a lo\-we\-ring of the core bolometric luminosity that ranges from 
roughly 17\% at masses $M \sim 30M_{\odot}$ to 27\% in masses $M \sim 
3M_{\odot}$ \citep{1979ApJ...230..230C}. A much complicate relation must then 
exist between spectra, masses and stellar ages than treated in the present 
work \citep{1985MNRAS.213..519C,1986serd.book.....Z,1987pbes.coll...68Z, 
1988CRASM.306.1225Z,1988CRASM.306.1265Z,1990amml.conf..239Z,Z1992}.\par
\begin{table}[t]
\caption{Kinetic energy ratios and ${\cal E}$ ratios calculated for
different values of $p$}
\label{conv}
\begin{center}
\begin{tabular}{ccc}
\hline
\noalign{\smallskip}
$p$ & ${\cal K}(p)/{\cal K}(\!p\!\!=\!\!0)$ & ${\cal E}(p)$  \\
\noalign{\smallskip}
\hline
\noalign{\smallskip}
 0.4 & 1.52 & 0.018 \\
 0.7 & 2.17 & 0.026 \\
 1.0 & 3.26 & 0.038 \\
\noalign{\smallskip}
\hline
\end{tabular}
\end{center}
\end{table}

\section{Conclusions}

 In this paper we have studied a sample of 97 field Be stars, most of which
are at distances $d <$ 500 pc from the Sun, so that they can be considered
more or less homogeneous regarding their initial metallicity. All these stars
were observed in the BCD spectrophotometric system to have photospheric 
spectral signatures as much as possible free of CE emission/absorption 
perturbations. The apparent fundamental parameters derived from the observed 
BCD ($\lambda_1,D$) quantities, i.e. parameters reflecting the average 
rotationally-perturbed photosphere shown by the projected stellar hemisphere 
towards the observer, were translated into $pnrc$ and $averaged$ fundamental 
parameters. ``$pnrc$" is the acronym for {\it parent non-rotating 
counterparts}, or parameters that correspond to homologous non-rotating stars. 
The $averaged$ fundamental parameters correspond to averages over the whole 
stellar surface. We have assumed that the studied Be stars rotate with an 
angular velocity ratio \omc = 0.88 \citep{FZHF2005}. Note the difference 
between $averaged$ and $apparent$, the last representing a sort of average 
spectrum emitted by the `observed' stellar hemisphere. The $averaged$ 
parameters are the only quantities that can be used to interpolate stellar 
masses and ages in the evolutionary tracks.\par
 The present contribution represents one of the first attempts to derive 
stellar masses and ages of Be stars by using simultaneously model atmospheres
and evolutionary tracks both calculated for rotating objects. According to 
the statistical average of true rotational velocities $V$ of dwarf Be stars, 
the evolutionary models used are for ZAMS equatorial rotational velocity $V_o 
=$ 300 km~s$^{-1}$ in all masses. For all stars we derived the mass and 
stellar ages $\tau$ normalized to the respective time that each rotating star
can spend in the main sequence phase $\tau_{\rm MS}$. As a consequence of 
effects of the rapid rotation described by the models used, we obtained a 
trend of points in the ($\tau/\tau_{\rm MS},M/M_{\odot}$) diagram, which 
implies that:\par
 a) there are Be stars spread over the whole age interval $0 \la \tau
/\tau_{\rm MS} \la 1$ in the main sequence evolutionary phase;\par
 b) in massive stars the Be phenomenon tends to be present at lower $\tau
/\tau_{\rm MS}$ age ratios than in the less massive stars.\par
 The lack of massive stars in the upper MS can be due to loss of angular 
momentum through mass-loss which produces a strong decrease of the stellar 
surface $\Omega/\Omega_{\rm c}$ ratio. On the other hand, the increase of the 
number of less massive Be stars in the upper MS can be explained in terms of 
angular momentum transport from the core to the surface carried by the 
meridional circulation, which has longer time scales the lower the stellar
mass.\par
 Arguments based on the distribution of the total angular momentum of dwarf
stars against mass reveal that the massive stars may start evolving from the 
ZAMS with a slightly higher $V_o/V_c$ than the less massive ones.\par
 
\begin{acknowledgements}
JZ warmly thanks Drs. A. Maeder, G. Meynet and J.P. Zahn for suggestions 
as well as discussions with Drs A.M. Hubert, M. Floquet and and J. Fabregat. 
YF acknowledges funding from the Belgian 'Diensten van de Eerste Minister - 
Federale Diensten voor We\-ten\-schap\-pelij\-ke, Technische en Culturele 
Aangelegenheden' (Research project MO/33/007). We are grateful to the unknown 
referee for valuable comments.\par 
 
\end{acknowledgements}

\bibliographystyle{aa}

\begin{thebibliography}{63}
\expandafter\ifx\csname natexlab\endcsname\relax\def\natexlab#1{#1}\fi


\bibitem[{{Baillet} {et~al.}(1973)}]{BCD1973} 
{Baillet}, A., {Chalonge}, D., \& {Divan}, L. 1973, Nouv. Rev. Optique, 4, 
151

\bibitem[{{Beech} \& {Mitalas}(1994)}]{1994ApJS...95..517B}
{Beech}, M., \& {Mitalas}, R. 1994, {\apjs}, 95, 517

\bibitem[{{Behrend} \& {Maeder}(2001)}]{2001A&A...373..190B}
{Behrend}, R., \& {Maeder}, A., {\aap}, 373, 190

\bibitem[{{Bodenheimer}(1971)}]{1971ApJ...167..153B}
{Bodenheimer}, P. 1971, \apj, 167, 153

\bibitem[{{Chalonge} \& {Divan}(1952)}]{1952AnAp...15..201C}
{Chalonge}, D. \& {Divan}, L. 1952, Annales d'Astrophysique, 15, 201

\bibitem[{{Chauville} {et~al.}(2001){Chauville}, {Zorec}, {Ballereau},
  {Morrell}, {Cidale}, \& {Garcia}}]{2001A&A...378..861C}
{Chauville}, J., {Zorec}, J., {Ballereau}, D., {et~al.} 2001, \aap, 378, 861

\bibitem[{{Cidale} {et~al.}(2000){Cidale}, {Zorec}, \&
  {Morrell}}]{2000ASPC..214...87C}
{Cidale}, L., {Zorec}, J., \& {Morrell}, N. 2000, ASPC, 214, 87

\bibitem[{{Cidale} {et~al.}(2001){Cidale}, {Zorec}, \&
  {Tringaniello}}]{2001A&A...368..160C}
{Cidale}, L., {Zorec}, J., \& {Tringaniello}, L. 2001, \aap, 368, 160

\bibitem[{{Claret}(1998)}]{1998A&A...335..647C}
{Claret}, A. 1998, \aap, 335, 647

\bibitem[{{Clement}(1979)}]{1979ApJ...230..230C}
{Clement}, M.~J. 1979, \apj, 230, 230

\bibitem[{{Coe}(2000)}]{2000bpet.conf..656C}
{Coe}, M.~J. 2000, in ASP Conf. Ser. 214: IAU Colloq. 175: The Be Phenomenon in
  Early-Type Stars, 656

\bibitem[{{Collins} \& {Smith}(1985)}]{1985MNRAS.213..519C}
{Collins}, G.~W. \& {Smith}, R.~C. 1985, \mnras, 213, 519

\bibitem[{{Collins} \& {Sonneborn}(1977)}]{1977ApJS...34...41C}
{Collins}, G.~W. \& {Sonneborn}, G.~H. 1977, \apjs, 34, 41

\bibitem[{{Collins} \& {Truax}(1995)}]{1995ApJ...439..860C}
{Collins}, G.~W. \& {Truax}, R.~J. 1995, \apj, 439, 860

\bibitem[{{Collins} {et~al.}(1991){Collins}, {Truax}, \&
  {Cranmer}}]{1991ApJS...77..541C}
{Collins}, G.~W., {Truax}, R.~J., \& {Cranmer}, S.~R. 1991, \apjs, 77, 541

\bibitem[{{Crampin} \& {Hoyle}(1960)}]{1960MNRAS.120...33C}
{Crampin}, J. \& {Hoyle}, F. 1960, \mnras, 120, 33

\bibitem[{{Denissenkov} {et~al.}(1999){Denissenkov}, {Ivanova}, \&
  {Weiss}}]{1999A&A...341..181D}
{Denissenkov}, P.~A., {Ivanova}, N.~S., \& {Weiss}, A. 1999, \aap, 341, 181

\bibitem[{{Deupree}(1998)}]{1998ApJ...499..340D}
{Deupree}, R.~G. 1998, \apj, 499, 340

\bibitem[{{Deupree}(2000)}]{2000ApJ...543..395D}
{Deupree}, R.~G. 2000, \apj, 543, 395

\bibitem[{{Deupree}(2001)}]{2001ApJ...552..268D}
{Deupree}, R.~G. 2001, \apj, 552, 268

\bibitem[Divan \& Zorec(1982)]{DZ1982} 
{Divan}, L., \& {Zorec}, J. 1982, ESA SP-177: The Scientific Aspects of the 
Hipparcos Space Astrometry Mission, 101

\bibitem[{{Endal}(1982)}]{1982IAUS...98..299E}
{Endal}, A.~S. 1982, in IAU Symp. 98: Be Stars, 299--302

\bibitem[{{Endal} \& {Sofia}(1979)}]{1979ApJ...232..531E}
{Endal}, A.~S. \& {Sofia}, S. 1979, \apj, 232, 531

\bibitem[{{Endal} \& {Sofia}(1981)}]{1981ApJ...243..625E}
{Endal}, A.~S. \& {Sofia}, S. 1981, \apj, 243, 625

\bibitem[{{Fabregat} \& {Torrej{\' o}n}(2000)}]{2000A&A...357..451F}
{Fabregat}, J. \& {Torrej{\' o}n}, J.~M. 2000, \aap, 357, 451

\bibitem[{{Feinstein}(1990)}]{1990RMxAA..21..373F}
{Feinstein}, A. 1990, Revista Mexicana de Astronomia y Astrofisica, vol.~21,
 21, 373

\bibitem[{{Fr\'emat} {et~al.}(2005)}]{FZHF2005}
{Fr\'emat}, Y., {Zorec}, J., {Hubert}, A.-M., \& {Floquet}, M. 2005, \aap,
(submitted)

\bibitem[{{Fujimoto}(1987)}]{1987A&A...176...53F}
{Fujimoto}, M.~Y. 1987, \aap, 176, 53

\bibitem[{{Gies}(2000)}]{2000bpet.conf..668G}
{Gies}, D.~R. 2000, in ASP Conf. Ser. 214: IAU Colloq. 175: The Be Phenomenon
  in Early-Type Stars, 668

\bibitem[{{Hardorp} \& {Strittmatter}(1970)}]{1970stro.coll...48H}
{Hardorp}, J. \& {Strittmatter}, P.~A. 1970, in IAU Colloq. 4: Stellar
  Rotation, 48

\bibitem[{{Harmanec}(1987)}]{1987pbes.coll..339H}
{Harmanec}, P. 1987, in IAU Colloq. 92: Physics of Be Stars, 339--355

\bibitem[{{Heger} \& {Langer}(2000)}]{2000ApJ...544.1016H}
{Heger}, A. \& {Langer}, N. 2000, \apj, 544, 1016

\bibitem[{{Hubert} {et~al.}(2000){Hubert}, {Floquet}, \&
  {Zorec}}]{2000bpet.conf..348H}
{Hubert}, A.~M., {Floquet}, M., \& {Zorec}, J. 2000, in ASP Conf. Ser. 214: IAU
  Colloq. 175: The Be Phenomenon in Early-Type Stars, 348

\bibitem[{{Hubert-Delplace} {et~al.}(1982){Hubert-Delplace}, {Hubert},
  {Chambon}, \& {Jaschek}}]{1982IAUS...98..125H}
{Hubert-Delplace}, A.~M., {Hubert}, H., {Chambon}, M.~T., \& {Jaschek}, M.
  1982, in IAU Symp. 98: Be Stars, 125--128

\bibitem[{{Jaschek} {et~al.}(1980){Jaschek}, {Jaschek}, {Hubert-Delplace}, \&
  {Hubert}}]{1980A&AS...42..103J}
{Jaschek}, M., {Jaschek}, C., {Hubert-Delplace}, A.-M., \& {Hubert}, H. 1980,
  \aaps, 42, 103

\bibitem[{{Jaschek} {et~al.}(1981){Jaschek}, {Slettebak}, \& 
{Jaschek}}]{JSJ1981} {Jaschek}, M., {Slettebak}, A., \& {Jaschek}, C. (1981),
Be Stars Newsl. No. 4, 9
 
\bibitem[{{Kawaler}(1987)}]{1987PASP...99.1322K}
{Kawaler}, S.~D. 1987, \pasp, 99, 1322

\bibitem[{{Maeder} \& {Behrend}(2002)}]{2002ASPC..267..179M}
{Maeder}, A., \& {Behrend}, R. 2002, ASPC, 267, 179

\bibitem[{{Maeder} {et~al.}(1999){Maeder}, {Grebel}, \&
  {Mermilliod}}]{1999A&A...346..459M}
{Maeder}, A., {Grebel}, E.~K., \& {Mermilliod}, J. 1999, \aap, 346, 459

\bibitem[{{Maeder} \& {Meynet}(2001)}]{2001A&A...373..555M}
{Maeder}, A. \& {Meynet}, G. 2001, \aap, 373, 555

\bibitem[{{Maeder} \& {Meynet}(2003)}]{2003A&A...411..543M}
{Maeder}, A. \& {Meynet}, G. 2003, \aap, 411, 543

\bibitem[{{Maeder} \& {Peytremann}(1970)}]{1970A&A.....7..120M}
{Maeder}, A. \& {Peytremann}, E. 1970, \aap, 7, 120

\bibitem[{{Maeder} \& {Peytremann}(1972)}]{1972A&A....21..279M}
{Maeder}, A. \& {Peytremann}, E. 1972, \aap, 21, 279

\bibitem[{{Massa}(1975)}]{1975PASP...87..777M}
{Massa}, D. 1975, \pasp, 87, 777

\bibitem[{{Mercer-Smith} {et~al.}(1999){Mercer-Smith}, {Cameron}, \& 
{Epstein}}]{1984ApJ...279..363M}
{Mercer-Smith}, J.~A. and {Cameron}, A.~G.~W., \& {Epstein}, R.~I. 1984,
{\apj}, 279, 363

\bibitem[{{Mermilliod}(1982)}]{1982A&A...109...48M}
{Mermilliod}, J.~C. 1982, \aap, 109, 48

\bibitem[{{Meynet} \& {Maeder}(2000)}]{2000A&A...361..101M}
{Meynet}, G. \& {Maeder}, A. 2000, \aap, 361, 101 (MM2000)

\bibitem[{{Meynet} \& {Maeder}(2002)}]{2002A&A...390..561M}
{Meynet}, G. \& {Maeder}, A. 2002, \aap, 390, 561

\bibitem[{{Moss} \& {Smith}(1982)}]{1982RPPh...44..831M}
{Moss}, D. \& {Smith}, R.~C. 1982, Reports of Progress in Physics, 44, 831

\bibitem[{{Moujtahid} {et~al.}(1999){Moujtahid}, {Zorec}, \&
  {Hubert}}]{1999A&A...349..151M}
{Moujtahid}, A., {Zorec}, J., \& {Hubert}, A.~M. 1999, \aap, 349, 151

\bibitem[{{Moujtahid} {et~al.}(1998){Moujtahid}, {Zorec}, {Hubert}, {Garcia},
  \& {Burki}}]{1998A&AS..129..289M}
{Moujtahid}, A., {Zorec}, J., {Hubert}, A.~M., {Garcia}, A., \& {Burki}, G.
  1998, \aaps, 129, 289

\bibitem[{{Owocki} (2004)}]{O2004}
{Owocki}, S.~P. 2004, in Stellar Rotation, IAU Symp. 215,
(eds.) A. Maeder \& Ph. Eenens, ASP Conf. Ser. \& IAU Publ., p. 515

\bibitem[{{Packet}(1981)}]{1981A&A...102...17P}
{Packet}, W. 1981, \aap, 102, 17

\bibitem[{{Palla} \& {Stahler}(1993)}]{1993ApJ...418..414P}
{Palla}, F., \& {Stahler}, S.~W. 1993, {\apj}, 418, 414

\bibitem[{{Pols} {et~al.}(1991){Pols}, {Cote}, {Waters}, \&
  {Heise}}]{1991A&A...241..419P}
{Pols}, O.~R., {Cote}, J., {Waters}, L.~B.~F.~M., \& {Heise}, J. 1991, \aap,
  241, 419

\bibitem[{{Sackmann}(1970)}]{1970A&A.....8...76S}
{Sackmann}, I.~J. 1970, \aap, 8, 76

\bibitem[{{Schaller} {et~al.}(1992){Schaller}, {Schaerer}, {Meynet}, \&
  {Maeder}}]{1992A&AS...96..269S}
{Schaller}, G., {Schaerer}, D., {Meynet}, G., \& {Maeder}, A. 1992, \aaps, 96,
  269

\bibitem[{{Schild} \& {Romanishin}(1976)}]{1976ApJ...204..493S}
{Schild}, R. \& {Romanishin}, W. 1976, \apj, 204, 493

\bibitem[{{Schmidt-Kaler}(1964)}]{1964VeBon..70....1S}
{Schmidt-Kaler}, T. 1964, Veroeffentlichungen des Astronomisches Institute der
  Universitaet Bonn, 70, 1

\bibitem[{{Slettebak}(1979)}]{1979SSRv...23..541S}
{Slettebak}, A. 1979, Space Science Reviews, 23, 541

\bibitem[{{Slettebak}(1985)}]{1985ApJS...59..769S}
{Slettebak}, A. 1985, \apjs, 59, 769

\bibitem[{{Slettebak} {et~al.}(1980){Slettebak}, {Kuzma}, \&
  {Collins}}]{1980ApJ...242..171S}
{Slettebak}, A., {Kuzma}, T.~J., \& {Collins}, G.~W. 1980, \apj, 242, 171

\bibitem[{{Spruit}(1999)}]{1999A&A...349..189S}
{Spruit}, H.~C. 1999, \aap, 349, 189

\bibitem[{{Spruit}(2002)}]{2002A&A...381..923S}
{Spruit}, H.~C. 2002, \aap, 381, 923

\bibitem[{{St{\c e}pie{\' n}}(2002)}]{2002A&A...383..218S}
{St{\c e}pie{\' n}}, K. 2002, \aap, 383, 218

\bibitem[{{Stoeckley}(1968)}]{1968MNRAS.140..141S}
{Stoeckley}, T.~R. 1968, \mnras, 140, 141

\bibitem[{{Stoeckley} \& {Mihalas}(1973)}]{Stoeckley1973}
{Stoeckley}, T.~R. \& {Mihalas}, D. 1973, Limb darkening and rotation 
broadening of He\,{\sc i} and Mg\,{\sc ii} line profiles in early-type stars,
NCAR-TN/STR-84

\bibitem[{{Struve}(1931)}]{S1931}
{Struve}, O. 1931, \apj, 73, 94 

\bibitem[{{Talon} {et~al.}(1997){Talon}, {Zahn}, {Maeder}, \&
  {Meynet}}]{1997A&A...322..209T}
{Talon}, S., {Zahn}, J.-P., {Maeder}, A., \& {Meynet}, G. 1997, \aap, 322, 209

\bibitem[{{Tassoul}(1978)}]{1978trs..book.....T}
{Tassoul}, J. 1978, {Theory of rotating stars} (Princeton Series in
  Astrophysics, Princeton: University Press, 1978)

\bibitem[{{Terebey} {et~al.}(1984){Terebey}, {Shu}, \& 
{Cassen}}]{1984ApJ...286..529T}
{Terebey}, S., {Shu}, F.~H., \& {Cassen}, P. 1984, {\apj}, 286, 529

\bibitem[{{Townsend} {et~al.}(2004){Townsend}, {Owocki}, \&
  {Howarth}}]{2004MNRAS.350..189T}
{Townsend}, R.~H.~D., {Owocki}, S.~P., \& {Howarth}, I.~D. 2004, \mnras, 350,
  189

\bibitem[{{van Bever} \& {Vanbeveren}(1997)}]{1997A&A...322..116V}
{van Bever}, J. \& {Vanbeveren}, D. 1997, \aap, 322, 116

\bibitem[{{von Zeipel}(1924{\natexlab{a}})}]{1924MNRAS..84..665V}
{von Zeipel}, H. 1924{\natexlab{a}}, \mnras, 84, 665

\bibitem[{{von Zeipel}(1924{\natexlab{b}})}]{1924MNRAS..84..684V}
{von Zeipel}, H. 1924{\natexlab{b}}, \mnras, 84, 684

\bibitem[{{Wolff} {et~al.}(2004){Wolff}, {Strom}, \&
  {Hillenbrand}}]{2004ApJ...601..979W}
{Wolff}, S.~C., {Strom}, S.~E., \& {Hillenbrand}, L.~A. 2004, \apj, 601, 979

\bibitem[{{Zorec}(1986)}]{1986serd.book.....Z}
{Zorec}, J. 1986, {Structure et rotation differentielle dans le etoiles B 
avec et sans emission} (Th\`ese d'\'Etat Paris: Universite VII)

\bibitem[{{Zorec}(1992)}]{Z1992}
{Zorec}, J. 1992, {Hipparcos: Une nouvelle donne pour l'Astronomie} 
(Observatoire de la C\^ote d'Azur \& Soci\'et\'e fran\c caise des 
Sp\'ecialistes d'Astronomie), 407

\bibitem[{{Zorec}(2004)}]{Z2004}
{Zorec}, J. 2004, in Stellar Rotation, IAU Symp. 215,
(eds.) A. Maeder \& Ph. Eenens, ASP Conf. Ser. \& IAU Publ., p. 73

\bibitem[{{Zorec} \& {Briot}(1991)}]{1991A&A...245..150Z}
{Zorec}, J. \& {Briot}, D. 1991, \aap, 245, 150

\bibitem[{{Zorec} \& {Briot}(1997)}]{1997A&A...318..443Z}
{Zorec}, J. \& {Briot}, D. 1997, \aap, 318, 443

\bibitem[{{Zorec} {et~al.}(1987){Zorec}, {Divan}, {Mochkovitch}, \&
  {Garcia}}]{1987pbes.coll...68Z}
{Zorec}, J., {Divan}, L., {Mochkovitch}, R., \& {Garcia}, A. 1987, in IAU
  Colloq. 92: Physics of Be Stars, 68--70

\bibitem[{{Zorec} {et~al.}(2002){Zorec}, {Fr{\' e}mat}, {Hubert}, \&
  {Floquet}}]{2002rnpp.conf..244Z}
{Zorec}, J., {Fr{\' e}mat}, Y., {Hubert}, A.~M., \& {Floquet}, M. 2002, in ASP
  Conf. Ser. 259: IAU Colloq. 185: Radial and Nonradial Pulsationsn as Probes
  of Stellar Physics, 244--245

\bibitem[{{Zorec} {et~al.}(2004){Zorec}, {Levenhagen}, {Chauville},
  {Royer}, {Leister}, {Fr\'emat}, \& {Ballereau}}]{ZLCRLFB2004}
{Zorec}, J., {Levenhagen}, R., {Chauville}, J., {Royer}, F., {Leister}, N.V.,
{Fr\'emat}, Y., \& , {Ballereau}, D. 2004, in Stellar Rotation, IAU Symp. 215,
(eds.) A. Maeder \& Ph. Eenens, ASP Conf. Ser. \& IAU Publ., p. 89 

\bibitem[{{Zorec} {et~al.}(1988{\natexlab{a}}){Zorec}, {Mochkovitch}, \&
  {Divan}}]{1988CRASM.306.1265Z}
{Zorec}, J., {Mochkovitch}, R., \& {Divan}, L. 1988{\natexlab{a}}, Academie des
  Sciences Paris Comptes Rendus Serie Sciences Mathematiques, 306, 1265

\bibitem[{{Zorec} {et~al.}(1988{\natexlab{b}}){Zorec}, {Mochkovitch}, \&
  {Garcia}}]{1988CRASM.306.1225Z}
{Zorec}, J., {Mochkovitch}, R., \& {Garcia}, A. 1988{\natexlab{b}}, Academie
  des Sciences Paris Comptes Rendus Serie Sciences Mathematiques, 306, 1225

\bibitem[{{Zorec} {et~al.}(1990){Zorec}, {Mochkovitch}, \&
  {Garcia}}]{1990amml.conf..239Z}
{Zorec}, J., {Mochkovitch}, R.~A., \& {Garcia}, A. 1990, in NATO ASIC Proc.
  316: Angular Momentum and Mass Loss for Hot Stars, 239

\end{thebibliography}

\onecolumn

\begin{table}
\caption{Program stars, observed and derived parameters}
\label{tabdat1}
\begin{tabular}{ccc|ccc|cc|ccc|ccc}

\hline
\noalign{\smallskip}
 HD & $D$ & $\lambda_1$ & $T_{\rm eff}$ & $\log g$ & $V\!\sin i$ &
 $\overline{T_{\rm eff}}$ & $\log\overline{g}$ & $M/M_{\odot}$ & $\log\tau$ & 
 $\tau/\tau_{\rm MS}$ & $M/M_{\odot}$ & $\log\tau$ & $\tau/\tau_{\rm MS}$ \\
    & dex & \AA         &   K            & dex      & km/s  &
    K                     &        dex         &               &   $\tau$(yr)   & 
                      &               & $\tau$(yr) &                      \\
\hline
    &     &             &  \multicolumn{3}{c|}{Apparent}          &
\multicolumn{2}{c|}{Averaged} & \multicolumn{3}{c|}{$\Omega/\Omega_c = 0$}  &
\multicolumn{3}{c}{$\Omega/\Omega_c = 0.88$}  \\
\hline
   4180 & 0.269 & 41.2 & 14400 & 3.36 & 195 & 14800 & 3.48 &  5.7 & 7.85 & 1.01 &  6.0 & 7.87 & 0.96  \\
   5394 & 0.062 & 66.2 & 28670 & 4.20 & 432 & 30240 & 4.06 & 14.3 & 6.81 & 0.53 & 16.1 & 6.64 & 0.32  \\
  10144 & 0.241 & 39.9 & 15040 & 3.48 & 235 & 16090 & 3.47 &  6.2 & 7.76 & 1.02 &  6.7 & 7.74 & 0.96  \\
  10516 & 0.079 & 61.3 & 26540 & 4.20 & 440 & 28760 & 4.18 & 11.7 & 6.85 & 0.42 & 13.8 & 6.28 & 0.12  \\
  20336 & 0.179 & 59.4 & 18720 & 4.20 & 328 & 19620 & 4.10 &  6.4 & 7.46 & 0.56 &  7.1 & 7.20 & 0.31  \\
  22192 & 0.230 & 45.0 & 15670 & 3.69 & 275 & 16840 & 3.62 &  6.2 & 7.74 & 0.97 &  6.7 & 7.70 & 0.88  \\
  23016 & 0.377 & 43.1 & 12130 & 3.71 & 235 & 12790 & 3.74 &  4.0 & 8.16 & 0.91 &  4.3 & 8.13 & 0.79  \\
  23302 & 0.343 & 40.4 & 12460 & 3.44 & 170 & 13010 & 3.57 &  4.6 & 8.08 & 1.01 &  4.7 & 8.08 & 0.92  \\
  23480 & 0.299 & 44.6 & 13620 & 3.69 & 240 & 14190 & 3.69 &  4.8 & 7.97 & 0.93 &  5.0 & 7.97 & 0.83  \\
  23630 & 0.363 & 31.9 & 12260 & 3.05 & 140 & 12410 & 3.06 &  6.1 & 7.78 & 1.02 &  5.9 & 7.91 & 1.02  \\
  23862 & 0.381 & 53.1 & 12100 & 3.98 & 290 & 12890 & 4.03 &  3.4 & 8.20 & 0.66 &  3.8 & 8.01 & 0.45  \\
  24534 & 0.062 & 59.9 & 28120 & 4.20 & 293 & 28140 & 3.96 & 14.3 & 6.88 & 0.62 & 14.3 & 6.89 & 0.50  \\
  25940 & 0.228 & 48.2 & 16180 & 3.82 & 197 & 16720 & 3.79 &  6.0 & 7.73 & 0.89 &  6.1 & 7.72 & 0.73  \\
  28497 & 0.063 & 55.9 & 28090 & 4.08 & 342 & 28970 & 4.13 & 13.3 & 6.79 & 0.45 & 14.3 & 6.52 & 0.21  \\
  30076 & 0.129 & 46.8 & 21410 & 3.85 & 213 & 21960 & 3.87 &  9.1 & 7.32 & 0.82 &  9.3 & 7.29 & 0.65  \\
  32343 & 0.235 & 59.5 & 16100 & 4.11 &  95 & 15460 & 4.02 &  5.2 & 7.74 & 0.63 &  5.0 & 7.73 & 0.47  \\
  35411 & 0.126 & 98.3 & 22690 & 4.19 & 170 & 21970 & 3.88 &  9.7 & 7.23 & 0.73 &  9.3 & 7.29 & 0.64  \\
  35439 & 0.134 & 52.5 & 22300 & 3.98 & 263 & 22800 & 3.98 &  9.1 & 7.25 & 0.69 &  9.5 & 7.16 & 0.49  \\
  36576 & 0.122 & 53.7 & 22880 & 4.02 & 265 & 23660 & 3.90 & 10.1 & 7.22 & 0.76 & 10.4 & 7.16 & 0.63  \\
  37202 & 0.155 & 44.2 & 19610 & 3.74 & 310 & 20050 & 3.68 &  8.5 & 7.42 & 0.89 &  8.6 & 7.46 & 0.81  \\
  37490 & 0.163 & 48.3 & 20100 & 3.51 & 172 & 19000 & 3.44 &  8.4 & 7.45 & 0.91 &  8.1 & 7.53 & 0.83  \\
  37657 & 0.195 & 48.0 & 17530 & 3.76 & 198 & 17870 & 3.77 &  6.8 & 7.60 & 0.89 &  6.9 & 7.62 & 0.76  \\
  37795 & 0.326 & 43.0 & 12810 & 3.64 & 180 & 13140 & 3.64 &  4.6 & 8.05 & 0.96 &  4.6 & 8.08 & 0.87  \\
  37967 & 0.239 & 55.0 & 15900 & 4.02 & 210 & 15920 & 3.93 &  5.3 & 7.78 & 0.74 &  5.4 & 7.75 & 0.59  \\
  38010 & 0.069 & 58.0 & 27480 & 4.19 & 370 & 28470 & 4.28 & 11.8 & 6.37 & 0.14 & 13.0 & 5.35 & 0.01  \\
  40978 & 0.180 & 61.5 & 18430 & 3.77 & 200 & 19020 & 3.81 &  7.3 & 7.53 & 0.86 &  7.5 & 7.52 & 0.71  \\
  41335 & 0.129 & 51.0 & 21800 & 3.92 & 358 & 23650 & 4.02 &  8.7 & 7.28 & 0.68 & 10.1 & 7.07 & 0.44  \\
  42545 & 0.238 & 56.0 & 15600 & 4.08 & 285 & 16110 & 3.97 &  5.1 & 7.81 & 0.72 &  5.4 & 7.71 & 0.54  \\
  44458 & 0.086 & 44.0 & 23870 & 3.55 & 242 & 24570 & 3.63 & 12.9 & 7.11 & 0.91 & 12.9 & 7.16 & 0.83  \\
  45314 & 0.040 & 55.2 & 30630 & 4.04 & 285 & 30620 & 4.08 & 16.2 & 6.65 & 0.43 & 16.5 & 6.58 & 0.28  \\
  45542 & 0.300 & 46.0 & 13480 & 3.68 & 217 & 14100 & 3.73 &  4.7 & 7.99 & 0.92 &  4.9 & 7.99 & 0.80  \\
  45725 & 0.192 & 55.0 & 18070 & 4.06 & 330 & 18940 & 3.95 &  6.5 & 7.58 & 0.73 &  7.0 & 7.47 & 0.56  \\
  45910 & 0.078 & 32.2 & 20660 & 2.74 & 240 & 22320 & 2.77 & 22.1 & 6.84 & 1.01 & 23.3 & 6.92 & 1.01  \\
  45995 & 0.120 & 51.7 & 22440 & 3.94 & 255 & 22570 & 3.82 &  9.9 & 7.25 & 0.79 &  9.9 & 7.27 & 0.68  \\
  47054 & 0.361 & 43.1 & 12490 & 3.68 & 219 & 13090 & 3.70 &  4.3 & 8.10 & 0.94 &  4.5 & 8.09 & 0.83  \\
  50013 & 0.092 & 87.9 & 26380 & 4.20 & 243 & 25800 & 4.12 & 11.5 & 6.81 & 0.37 & 11.4 & 6.73 & 0.25  \\
  50083 & 0.124 & 48.0 & 21930 & 3.85 & 170 & 22500 & 3.77 & 10.1 & 7.26 & 0.87 & 10.1 & 7.28 & 0.73  \\
  53974 & 0.054 & 51.0 & 28910 & 3.84 & 100 & 27810 & 3.69 & 18.1 & 6.85 & 0.80 & 16.1 & 7.01 & 0.77  \\
  56014 & 0.169 & 37.8 & 18540 & 3.29 & 280 & 20330 & 3.34 &  9.8 & 7.36 & 1.02 & 10.7 & 7.36 & 0.98  \\
  56139 & 0.190 & 55.6 & 18110 & 4.05 &  85 & 17360 & 3.87 &  6.6 & 7.58 & 0.77 &  6.3 & 7.64 & 0.66  \\
  57219 & 0.218 & 46.5 & 16500 & 3.81 &  80 & 15850 & 3.71 &  6.2 & 7.70 & 0.89 &  5.9 & 7.80 & 0.81  \\
  58050 & 0.162 & 60.2 & 20100 & 4.20 & 130 & 19360 & 4.02 &  7.3 & 7.37 & 0.59 &  7.1 & 7.36 & 0.45  \\
  58343 & 0.230 & 48.0 & 16250 & 3.79 &  43 & 15520 & 3.73 &  6.1 & 7.72 & 0.89 &  5.7 & 7.84 & 0.80  \\
  58715 & 0.400 & 50.9 & 11740 & 3.88 & 230 & 12060 & 3.94 &  3.5 & 8.26 & 0.75 &  3.6 & 8.18 & 0.58  \\
  60848 & 0.051 & 71.1 & 29930 & 4.20 & 247 & 30760 & 4.33 & 14.2 & 6.22 & 0.13 & 14.9 & ...  & 0.00  \\
  63462 & 0.050 & 78.0 & 30020 & 4.20 & 435 & 32020 & 4.04 & 16.4 & 6.78 & 0.58 & 18.5 & 6.60 & 0.35  \\
  65875 & 0.142 & 54.0 & 21350 & 4.04 & 153 & 20670 & 3.92 &  8.5 & 7.32 & 0.69 &  8.2 & 7.36 & 0.58  \\
  68980 & 0.072 & 57.7 & 26800 & 4.20 & 145 & 25840 & 4.08 & 12.0 & 6.84 & 0.44 & 11.6 & 6.82 & 0.34  \\
  83953 & 0.253 & 49.7 & 14800 & 3.78 & 260 & 15650 & 3.81 &  5.1 & 7.88 & 0.86 &  5.5 & 7.82 & 0.72  \\
  91120 & 0.438 & 50.0 & 10780 & 3.95 & 307 & 11950 & 3.92 &  3.2 & 8.38 & 0.81 &  3.6 & 8.21 & 0.61  \\
  91465 & 0.193 & 43.0 & 17200 & 3.63 & 266 & 18630 & 3.66 &  7.1 & 7.60 & 0.96 &  7.7 & 7.56 & 0.82  \\
 109387 & 0.288 & 45.0 & 14174 & 3.59 & 200 & 14670 & 3.69 &  5.2 & 7.91 & 0.93 &  5.3 & 7.92 & 0.83  \\
 110432 & 0.150 & 80.0 & 20350 & 3.92 & 400 & 22510 & 3.90 &  8.1 & 7.40 & 0.76 &  9.6 & 7.24 & 0.61  \\
 112091 & 0.151 & 63.0 & 21190 & 4.05 & 210 & 21600 & 4.27 &  7.5 & 7.04 & 0.28 &  7.9 & 6.21 & 0.04  \\
 120324 & 0.112 & 54.4 & 23130 & 4.04 & 159 & 22410 & 3.95 &  9.6 & 7.18 & 0.65 &  9.3 & 7.22 & 0.55  \\
 120991 & 0.106 & 53.3 & 23320 & 3.95 &  70 & 22330 & 3.82 & 10.5 & 7.19 & 0.76 &  9.8 & 7.29 & 0.69  \\
 124367 & 0.222 & 57.8 & 16460 & 4.04 & 295 & 17060 & 4.08 &  5.2 & 7.68 & 0.57 &  5.7 & 7.48 & 0.36  \\
 127972 & 0.165 & 51.4 & 20050 & 3.94 & 310 & 20730 & 3.92 &  7.6 & 7.41 & 0.71 &  8.2 & 7.35 & 0.57  \\
\noalign{\smallskip}
\hline
\end{tabular}
\end{table}

\begin{table}
\caption{Program stars, observed and derived parameters}
\label{tabdat1}
\begin{tabular}{ccc|ccc|cc|ccc|ccc}

\hline
\noalign{\smallskip}
 HD & $D$ & $\lambda_1$ & $T_{\rm eff}$ & $\log g$ & $V\!\sin i$ &
 $\overline{T_{\rm eff}}$ & $\log\overline{g}$ & $M/M_{\odot}$ & $\log\tau$ & 
 $\tau/\tau_{\rm MS}$ & $M/M_{\odot}$ & $\log\tau$ & $\tau/\tau_{\rm MS}$ \\
    & dex & \AA         &   K            & dex      & km/s  &
    K                     &        dex         &               &   $\tau$(yr)   & 
                      &               & $\tau$(yr) &                      \\
\hline
    &     &             &  \multicolumn{3}{c|}{Apparent}          &
\multicolumn{2}{c|}{Averaged} & \multicolumn{3}{c|}{$\Omega/\Omega_c = 0$}  &
\multicolumn{3}{c}{$\Omega/\Omega_c = 0.88$}  \\
\hline
 131492 & 0.145 & 57.8 & 21120 & 4.09 & 185 & 20560 & 3.99 &  8.0 & 7.31 & 0.68 &  7.9 & 7.31 & 0.48  \\
 135734 & 0.356 & 54.5 & 12580 & 4.06 & 278 & 13290 & 4.06 &  3.6 & 8.14 & 0.66 &  4.0 & 7.94 & 0.42  \\
 137387 & 0.134 & 69.1 & 21980 & 4.20 & 250 & 21949 & 4.13 &  8.2 & 7.15 & 0.44 &  8.5 & 6.96 & 0.24  \\
 138749 & 0.278 & 52.6 & 14440 & 3.84 & 340 & 15969 & 3.96 &  4.6 & 7.93 & 0.73 &  5.4 & 7.73 & 0.55  \\
 142184 & 0.140 & 65.6 & 21540 & 4.20 & 340 & 22790 & 4.30 &  7.6 & 6.93 & 0.23 &  8.6 & 5.68 & 0.01  \\
 142926 & 0.383 & 54.0 & 12060 & 3.98 & 336 & 13320 & 4.08 &  3.4 & 8.20 & 0.63 &  4.0 & 7.89 & 0.37  \\
 142983 & 0.200 & 48.0 & 17790 & 3.80 & 390 & 20240 & 3.93 &  6.4 & 7.60 & 0.77 &  7.9 & 7.38 & 0.56  \\
 148184 & 0.046 & 66.3 & 30700 & 4.20 & 144 & 29590 & 4.14 & 15.6 & 6.47 & 0.27 & 14.9 & 6.42 & 0.17  \\
 149438 & 0.061 & 51.5 & 28600 & 3.83 &  15 & 27310 & 3.85 & 15.6 & 6.88 & 0.69 & 14.1 & 7.00 & 0.64  \\
 149757 & 0.068 & 60.2 & 27810 & 4.20 & 348 & 28610 & 4.01 & 13.8 & 6.89 & 0.60 & 14.5 & 6.81 & 0.42  \\
 157042 & 0.129 & 57.8 & 22050 & 4.19 & 340 & 22560 & 4.11 &  8.3 & 7.16 & 0.46 &  9.0 & 6.97 & 0.29  \\
 158427 & 0.201 & 58.6 & 17360 & 4.13 & 290 & 17780 & 4.04 &  5.8 & 7.61 & 0.60 &  6.2 & 7.48 & 0.43  \\
 162732 & 0.312 & 49.3 & 13070 & 3.81 & 310 & 14120 & 3.80 &  4.3 & 8.07 & 0.86 &  4.8 & 7.98 & 0.74  \\
 164284 & 0.141 & 62.6 & 21650 & 4.11 & 280 & 21830 & 3.98 &  8.2 & 7.23 & 0.52 &  8.5 & 7.08 & 0.33  \\
 173948 & 0.123 & 46.8 & 22250 & 3.83 & 140 & 21550 & 3.59 & 11.2 & 7.21 & 0.91 & 10.4 & 7.33 & 0.85  \\
 174237 & 0.198 & 56.0 & 17450 & 4.07 & 163 & 16960 & 3.83 &  6.4 & 7.63 & 0.81 &  6.2 & 7.69 & 0.70  \\
 178175 & 0.157 & 37.6 & 18940 & 3.49 & 105 & 18290 & 3.50 &  8.7 & 7.44 & 0.99 &  8.2 & 7.55 & 0.91  \\
 183656 & 0.288 & 35.0 & 12630 & 3.29 & 275 & 14110 & 3.30 &  5.3 & 7.92 & 1.02 &  5.9 & 7.90 & 1.02  \\
 184915 & 0.060 & 47.4 & 27830 & 3.60 & 229 & 28190 & 3.67 & 17.5 & 6.90 & 0.84 & 16.9 & 6.99 & 0.77  \\
 187811 & 0.190 & 53.0 & 17790 & 3.93 & 245 & 18600 & 4.00 &  6.3 & 7.60 & 0.72 &  6.7 & 7.45 & 0.48  \\
 192044 & 0.319 & 44.4 & 12940 & 3.66 & 245 & 13770 & 3.67 &  4.6 & 8.03 & 0.95 &  4.9 & 8.01 & 0.85  \\
 194335 & 0.133 & 70.9 & 22200 & 4.20 & 360 & 22970 & 4.23 &  8.1 & 6.90 & 0.23 &  8.9 & 6.40 & 0.07  \\
 200120 & 0.124 & 65.7 & 22650 & 4.20 & 379 & 23870 & 4.10 &  8.9 & 7.16 & 0.53 & 10.0 & 6.90 & 0.29  \\
 202904 & 0.162 & 57.8 & 20470 & 4.06 & 185 & 20460 & 4.08 &  7.5 & 7.34 & 0.57 &  7.7 & 7.20 & 0.35  \\
 203374 & 0.055 & 67.5 & 29130 & 4.20 & 333 & 29830 & 3.88 & 16.6 & 6.85 & 0.71 & 16.9 & 6.87 & 0.58  \\
 205637 & 0.175 & 38.5 & 17880 & 3.19 & 225 & 19070 & 3.40 &  8.7 & 7.45 & 1.02 &  9.2 & 7.48 & 0.97  \\
 206773 & 0.050 & 65.3 & 30460 & 4.20 & 390 & 31500 & 4.16 & 15.6 & 6.59 & 0.35 & 17.0 & 6.21 & 0.12  \\
 209014 & 0.366 & 53.4 & 12360 & 3.87 & 320 & 13350 & 3.84 &  3.8 & 8.17 & 0.81 &  4.3 & 8.06 & 0.69  \\
 209409 & 0.333 & 48.0 & 12770 & 3.81 & 280 & 13960 & 3.87 &  4.1 & 8.11 & 0.85 &  4.6 & 7.98 & 0.66  \\
 209522 & 0.228 & 46.1 & 16260 & 3.80 & 275 & 17020 & 3.76 &  6.0 & 7.72 & 0.86 &  6.4 & 7.69 & 0.77  \\
 210129 & 0.300 & 44.4 & 13390 & 3.62 & 130 & 13770 & 3.70 &  4.9 & 7.98 & 0.96 &  4.8 & 8.02 & 0.83  \\
 212076 & 0.093 & 48.3 & 24290 & 3.81 &  98 & 23340 & 3.82 & 11.3 & 7.13 & 0.76 & 10.6 & 7.22 & 0.68  \\
 212571 & 0.076 & 57.0 & 27020 & 4.19 & 230 & 27770 & 4.13 & 12.6 & 6.90 & 0.54 & 13.1 & 6.57 & 0.21  \\
 214748 & 0.385 & 38.0 & 11999 & 3.45 & 200 & 12400 & 3.46 &  4.4 & 8.12 & 1.02 &  4.6 & 8.14 & 0.98  \\
 217050 & 0.186 & 40.4 & 17660 & 3.54 & 340 & 20180 & 3.59 &  7.7 & 7.55 & 0.98 &  9.2 & 7.43 & 0.88  \\
 217543 & 0.208 & 52.5 & 16990 & 3.85 & 330 & 18610 & 3.95 &  5.9 & 7.68 & 0.76 &  6.8 & 7.49 & 0.55  \\
 217675 & 0.281 & 38.3 & 14140 & 3.17 & 292 & 15800 & 3.34 &  6.0 & 7.78 & 1.02 &  6.8 & 7.74 & 1.01  \\
 217891 & 0.286 & 51.3 & 14050 & 3.84 &  95 & 13530 & 3.83 &  4.6 & 7.97 & 0.82 &  4.4 & 8.04 & 0.71  \\
 224686 & 0.419 & 44.5 & 11320 & 3.76 & 300 & 12260 & 3.75 &  3.6 & 8.28 & 0.89 &  4.0 & 8.20 & 0.78  \\

\noalign{\smallskip}
\hline
\end{tabular}
\end{table}

\end{document}